\documentclass[aps,11pt]{revtex4}
\usepackage{dcolumn}
\usepackage{graphicx}
\usepackage{amsmath}
\usepackage{amsfonts}
\usepackage{amssymb}
\usepackage{psfrag}
\usepackage{wrapfig}
\usepackage{subfigure}
\usepackage{makeidx}
\usepackage{bm}
\usepackage{epsf}
\usepackage{hyperref}
 \usepackage{float}
\usepackage{color}
\usepackage{cases}
\usepackage{booktabs} %绘制三线表
\usepackage{makecell} %表格内换行
\usepackage{multirow}
\usepackage{bm}%加粗 
\usepackage{enumitem}
\usepackage{array}

\makeatletter
\newcommand{\eq}[1]{Eq.(\ref{#1})}

\newcommand{\Rmnum}[1]{\uppercase\expandafter{\romannumeral #1}}
\usepackage{amsmath} 

\begin{document} 

\title{Quasinormal modes and greybody factors of magnetically charged de Sitter black holes probed by massless external fields in Einstein–Euler–Heisenberg gravity}

\author{Ming Zhang$^{1,2}$\footnote{Correspondence: zhangming@xaau.edu.cn}, Guo-Xin Chen$^{1}$, Lei Zhang$^{1}$,  Sheng-Yuan Li$^{3}$, \\ Xufen Zhang$^{3}$\footnote{Correspondence: xfzhangyzu@126.com}, and De-Cheng Zou$^{4}$\footnote{Correspondence: dczou@jxnu.edu.cn}}

\address{
$^{1}$Faculty of Science, Xihang University, Xi'an 710077 China \\
$^{2}$National Joint Engineering Research Center for Special Pump System Technology, Xihang University, Xi’an 710077, China\\
$^{3}$Center for Gravitation and Cosmology, College of Physical Science and Technology, Yangzhou University, Yangzhou 225009, China \\
$^{4}$College of Physics and Communication Electronics, Jiangxi Normal University, Nanchang 330022, China 
}

\date{\today}

\begin{abstract}
\indent

This paper investigates the perturbation dynamics of massless scalar and electromagnetic fields on magnetically charged de Sitter (dS) black holes within the framework of string-inspired Euler-Heisenberg (EH) gravity. 
We calculate the quasinormal frequencies (QNFs) and discuss the influences of black hole magnetic charge $Q_{\mathrm{m}}$, the cosmological constant $\Lambda$, coupling parameter $\epsilon$ and multipole number $l$ on QNFs, emphasizing the relationships between these parameters and quasinormal modes (QNMs) behavior. We find that the results obtained through the  asymptotic iteration method (AIM) are in good agreement with those  obtained by the WKB method. Importantly, the Bernstein spectral method is employed as a rigorous cross-check for QNFs in the $l=0$ scalar perturbation sector, where the WKB approximation is often unreliable.  The greybody factor (GFs) is calculated using WKB method. The effects of the parameters $Q_{\mathrm{m}}$ and $\epsilon$ on the greybody factor are also studied. 
\end{abstract}

%\keywords{}

\maketitle

\section{Introduction}
\label{intro}

The Euler-Heisenberg (EH) Lagrangian, introduced in 1936\cite{Heisenberg:1936nmg}, serves as a nonlinear extension of quantum electrodynamics (QED), offering a classical approximation that surpasses Maxwell's theory in strong-field regimes where vacuum polarization becomes significant.
In this framework, the vacuum is modeled as a dynamically polarizable medium, with polarization and magnetization arising from virtual charge clouds surrounding real charges and currents \cite{Obukhov:2002xa}. This approach not only provides a more accurate classical representation of QED under extreme conditions but also serves as a powerful tool for investigating nonlinear phenomena in astrophysics and cosmology. Leveraging these properties, the first EH black hole solution—a magnetically charged, anisotropic Reissner-Nordström-like configuration with dyon degrees of freedom—was derived in 1956 \cite{Yajima:2000kw}. Since then, attention has expanded toward electrically charged solutions \cite{Yajima:2000kw,Ruffini:2013hia}, rotating generalizations \cite{Breton:2019arv,Amaro:2022yew,Wu:2021pgf}, and extensions within modified gravity frameworks \cite{Guerrero:2020uhn,Nashed:2021ctg}. More recently, inspired by string theory and Lovelock gravity,  Ref. \cite{Bakopoulos:2024hah} proposed a coupling of the dilaton field to EH electrodynamics, thereby enriching the spectrum of possible black hole solutions. In the astrophysical context, studies have explored the motion of test particles and gravitational lensing in magnetically charged backgrounds \cite{Yasir:2025npe,Vachher:2024fxs}, as well as their shadow images \cite{Xu:2024gjs}. Further, Jiang et al. \cite{Jiang:2024njc} analyzed the properties of geometrically thin, optically thick accretion disks around such objects. Zhang et al.\cite{Zhang:2025xqt} investigates field perturbations on magnetically charged black holes in this theory, calculating QNFs, comparing AIM/WKB results, and computing greybody factors via WKB.

Quasinormal modes (QNMs) represent the characteristic oscillation spectra of spacetime perturbations, and they play a central role in black hole physics. For asymptotically de Sitter (dS) black holes, where a positive cosmological constant introduces a cosmological horizon, QNMs encode not only the intrinsic structure of the black hole but also the dynamical interplay between the event and cosmological horizons. These modes are of particular relevance in gravitational wave astronomy, gauge/gravity correspondence, and stability analyses. Cho et al. \cite{Cho:2009cj,Cho:2011sf} developed the asymptotic iteration method (AIM), a robust iterative approach for accurately computing quasinormal frequencies (QNFs), which we will employ here. In parallel, we also use the WKB method \cite{Iyer:1986np,Iyer:1986nq, Konoplya:2003ii,Kokkotas:1988fm}, a well-established semi-analytic technique, to obtain independent QNF estimates and cross-check our results. To further ensure the precision of our results, we also incorporate the Bernstein spectral method\cite{Fortuna:2020obg} for a rigorous cross-check in the $l=0$ scalar perturbation sector, where the WKB approximation may be unreliable.

Another important quantity in the study of black hole perturbations is the greybody factor \cite{Konoplya:2019ppy,Konoplya:2011qq,Konoplya:2019hlu,Gogoi:2023fow,Lin:2024ubg}, which quantifies the modification of the radiation spectrum due to backscattering in the curved spacetime geometry. In the context of asymptotically dS spacetimes, where both the black hole and cosmological horizons contribute to scattering, the greybody factor plays a crucial role in connecting near-horizon emission processes to observable signals. The interplay between QNFs and greybody factors offers a unified description of wave dynamics: QNFs dictate the resonant ringdown behavior, while greybody factors determine the transmission efficiency of radiation to asymptotic observers \cite{Oshita:2023cjz}.

Motivated by these considerations, we investigate the QNFs and greybody factors of massless external fields propagating on magnetically charged dS black holes in the framework of Einstein–Euler–Heisenberg gravity. The paper is organized as follows. In Section \ref{sec2}, we briefly review the black hole solution with a positive cosmological constant. In Section \ref{sec3} we present the master equations for massless scalar and electromagnetic perturbations.  In Section \ref{sec4}, we compute the QNFs using the AIM, WKB  and   Bernstein  spectral   methods, examining the influences of black hole parameters, including the cosmological constant, on the spectra. In Section \ref{sec5},  We show the parametric dependence of the numerical results. In Section \ref{sec6}, we evaluate the greybody factors using the WKB method. Our conclusions and discussion are given in Section \ref{sec7}.

\section{Magnetically charged  de-sitter black holes}
\label{sec2}
Drawing on insights from string theory and Lovelock gravity, Bakopoulos \textit{et al.}~\cite{Bakopoulos:2024hah} recently proposed an Einstein--Maxwell--dilaton framework incorporating a nonlinear Euler--Heisenberg (EH) term coupled to the dilaton field. The action is given by~\cite{Bakopoulos:2024hah}
\begin{eqnarray}
S=\frac{1}{16\pi}\int  {\rm d}^4x\sqrt{-g}\Big[R-2\nabla^\mu\phi\nabla_\mu\phi-{\rm e}^{-2\phi}F^2
-f(\phi)\left(2\alpha F^\mu_{~\nu} F^\nu_{~\rho} F^\rho_{~\delta} F^\delta_{~\mu}-\beta F^4\right)
-\mathfrak{V}(\phi)\Big],
\label{action}
\end{eqnarray}
where $R$ is the Ricci scalar, $\phi$ denotes the dilaton field, and $f(\phi)$ is a scalar-dependent coupling function. We have defined
\[
F^2 = F_{\mu\nu}F^{\mu\nu}, \qquad
F^4 = F_{\mu\nu}F^{\mu\nu}F_{\rho\delta}F^{\rho\delta},
\]
with the electromagnetic field strength given by the usual expression
$F_{\mu\nu}=\partial_\mu \boldsymbol{A}_{\nu}-\partial_{\nu}\boldsymbol{A}_{\mu}$.
The coupling constants $\alpha$ and $\beta$ control the strength of the EH nonlinear corrections.
By varying the action \eqref{action} with respect to $g_{\mu\nu}$, $\phi$, and $\boldsymbol{A}_\mu$, we obtain the corresponding field equations:
\begin{eqnarray}
&&R_{\mu\nu}-\frac{1}{2}R g_{\mu\nu}+\frac{1}{2}g_{\mu\nu}\mathfrak{V(\phi)}-T_{\mu\nu}=0,\label{eqG}\\
&&\Box  \phi +\frac{1}{2}\mathrm{e}^{-2\phi}F^2-\frac{\mathrm{d}f(\phi)}{\mathrm{d}\phi}\left(\frac{\alpha}{2} F^\mu_{~\nu} F^\nu_{~\gamma} F^\gamma_{~\delta} F^\delta_{~\mu}-\frac{\beta}{4} F^4\right)-\frac{1}{4}\frac{\mathrm{d}\mathfrak{V(\phi)}}{\mathrm{d}\phi}=0,\label{eqKG}\\
&&{\partial_ \mu}\Big[\sqrt{-g}\left(\mathrm{e}^{-2\phi}F^{\mu\nu}+f(\phi)(4\alpha F^\mu_{~\kappa} F^\kappa_{~\lambda} F^{\nu\lambda}-2\beta F^2 F^{\mu\nu})\right)\Big]=0.\label{eqMaxwell}
\end{eqnarray}
The total energy–momentum tensor, incorporating contributions from the scalar and electromagnetic sectors, is
\begin{eqnarray}
T_{\mu\nu}=&&2{\partial _\mu}\phi{\partial_ \nu}\phi+g_{\mu\nu}{\partial }^\mu\phi {\partial }_\mu \phi-2\Bigg(\mathrm{e}^{-2\phi}(F^{\alpha}_\mu F_{\nu \alpha}-\frac{1}{4}g_{\mu\nu }F^2)\nonumber\\
&&+f(\phi)\Big( ( 4\alpha F^\alpha _\mu F^\beta_\nu F^\eta_\alpha F_{\beta\eta} -\frac{1}{2}\alpha g_{\mu\nu} F^\alpha_\beta F^\beta _\gamma F^\gamma _\delta F^\delta_\alpha -2\beta F^\xi_\mu F_{\nu\xi}F^2+\frac{1}{4}g_{\mu\nu}\beta F^4\Big)\Bigg),
\end{eqnarray}
For later convenience, we introduce the parameter
\begin{equation}
\epsilon \equiv \alpha-\beta ,
\end{equation}
which will appear explicitly in the black hole solution.
The dilaton potential~\cite{Gao:2004gc} is chosen as
\begin{eqnarray}
\mathfrak{V}(\phi)=\frac{1}{3}\Lambda \mathrm{e}^{-2\phi}
+\frac{1}{3}\Lambda \mathrm{e}^{2\phi}
+\frac{4\Lambda}{3},
\end{eqnarray}
where $\Lambda$ denotes the cosmological constant, and $\Lambda>0$ ensures that the spacetime is asymptotically de Sitter.
By choosing the coupling function~\cite{Bakopoulos:2024hah}
\begin{eqnarray}
 f(\phi)=-\Big[3\cosh(2\phi)+2\Big]
 \equiv -\frac{1}{2}\left(3\mathrm{e}^{-2\phi}+3\mathrm{e}^{2\phi}+4\right),
\label{phi}
\end{eqnarray}
It should be noted that the functional form of the coupling function $f(\phi)$ closely parallels that of the dilaton potential $\mathfrak{V}(\phi)$; both are constructed from Liouville-type exponential terms~\cite{2009PhRvD..80b4028C}.
the action~\eqref{action} admits a magnetically charged black hole solution with the metric~\cite{Bakopoulos:2024hah}
\begin{eqnarray}
\mathrm{d}s^2 &=& -A(r)\,\mathrm{d}t^2 + \frac{1}{B(r)}\,\mathrm{d}r^2 + r^2(\mathrm{d}\theta^2 + \sin^2\theta\,\mathrm{d}\varphi^2),
\label{metric}\\
A(r) &=& 1-\frac{4M^2}{Q_{\mathrm{m}}^2+\sqrt{Q_{\mathrm{m}}^4+4M^2r^2}}
-\frac{2\epsilon Q_{\mathrm{m}}^4}{r^6}
-\frac{1}{3}\Lambda r^2, \nonumber\\
B(r) &=& 1
-\frac{Q_{\mathrm{m}}^4+4M^2r^2}{r^2\!\left(Q_{\mathrm{m}}^2+\sqrt{Q_{\mathrm{m}}^4+4M^2r^2}\right)}
+\frac{Q_{\mathrm{m}}^4}{4M^2r^2}
-\frac{\epsilon Q_{\mathrm{m}}^4\!\left(Q_{\mathrm{m}}^4+4M^2r^2\right)}{2M^2r^8}
-\Lambda\!\left(\frac{r^2}{3}+\frac{Q_{\mathrm{m}}^4}{12M^2}\right), \nonumber\\
\phi(r) &=& -\frac{1}{2}\ln\!\left(
\frac{\sqrt{Q_{\mathrm{m}}^4+4M^2r^2}-Q_{\mathrm{m}}^2}
{\sqrt{Q_{\mathrm{m}}^4+4M^2r^2}+Q_{\mathrm{m}}^2}
\right), \quad
\boldsymbol{A}_\mu=(0,0,0,Q_{\mathrm{m}}\cos\theta).\label{solution}
\end{eqnarray}
where $M$ and $Q_{\mathrm{m}}$ represent the black hole mass and magnetic charge, respectively.

These solutions \eq{solution} are invariant under the following rescaling: $r/M\rightarrow r$, $Q_{\mathrm{m}} /M\rightarrow Q_{\mathrm{m}}$, and  $\epsilon/ M^2\rightarrow \epsilon$. Without loss of generality, we set $M=1$ in the following analysis and vary the parameters $\epsilon$, $\Lambda$ and $Q_{\mathrm{m}}$.
When $\epsilon = 0$ and $\Lambda = 0$, the metric \eqref{solution} reproduces the well-known  GMGHS or GHS black holes \cite{Gibbons:1987ps,Garfinkle:1990qj}. The latter further simplifies to the Schwarzschild solution for $Q_{\mathrm{m}} = 0$ and $\Lambda = 0$ , and both have been the subject of considerable investigation in the literature~\cite{Ferrari:2000ep,Chen:2004zr,Shu:2004fj,Karimov:2018whx}.
The solution \eqref{solution} describes a black hole with two horizon for $\epsilon=1$, while for $\epsilon=-1$, the black hole horizons can range from three to one. These imply the magnetically charged black hole possesses different horizon structures. Under variations of the parameter $\epsilon$ while maintaining a constant charge $Q_{\mathrm{m}}=0.3$, the horizon structure exhibits corresponding geometric modifications. 
\begin{enumerate}[label=(\roman*)] % 全局设置标签格式
    \item In the regime where $\epsilon\geq 0$, the spacetime structure permits only two horizons: the black hole event horizon and the cosmological horizon.\label{item:i}
    \item In the regime where $\epsilon<0$, he spacetime horizon structure undergoes significant transformations, see the right panels in Fig. \ref{B3}. For $\epsilon=-1$, representing weak coupling strength, three distinct horizons emerge: the Cauchy horizon $r_-$, black hole event horizon $r_{\mathrm{h}}$, and a cosmological horizon $r_{\mathrm{c}}$.  As the coupling intensifies to $\epsilon_{\mathrm{c}}\approx-292.357$, the system reaches a critical phase where the inner horizons merge into a single extremal black hole horizon $r_{\mathrm{h}}$, while the cosmological horizon $r_{\mathrm{c}}$ persists. In the strong coupling limit ($\epsilon=-1000$), only the cosmological horizon remains $(r=r_{\mathrm{c}})$, creating a causally connected region ($r < r_{\mathrm{c}}$ and $B(r) > 0$) that exposes the central singularity - a clear violation of cosmic censorship resulting from the dominant coupling effects. \label{item:ii}
\end{enumerate}
\begin{figure}[htb]
\centering
\subfigure[$\epsilon=-1$]
{\label{B2} %% label for first subfigure
\includegraphics[width=2in]{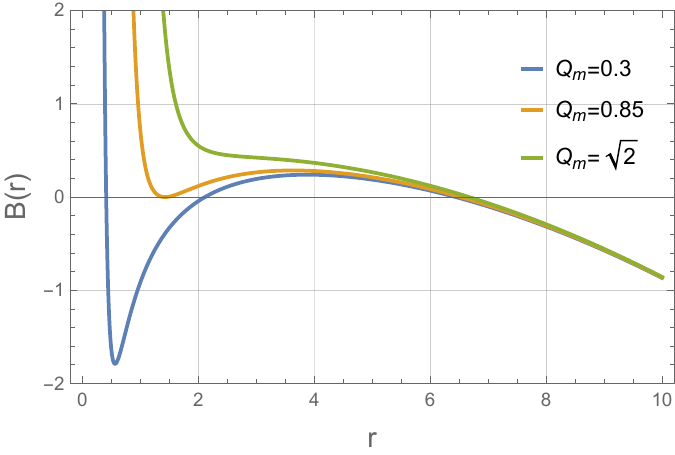}}
\hfill
\subfigure[$\epsilon=0$ ]
{\label{B0} %% label for first subfigure
\includegraphics[width=2in]{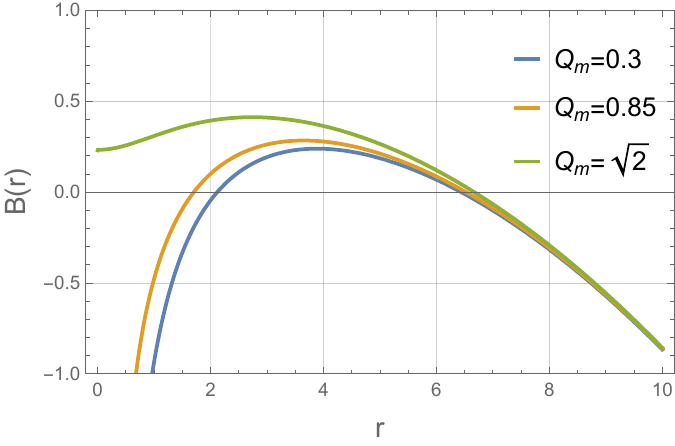}}
\hfill
\subfigure[$\epsilon=1$ ]
{\label{B1}%% label for first subfigure
\includegraphics[width=2in]{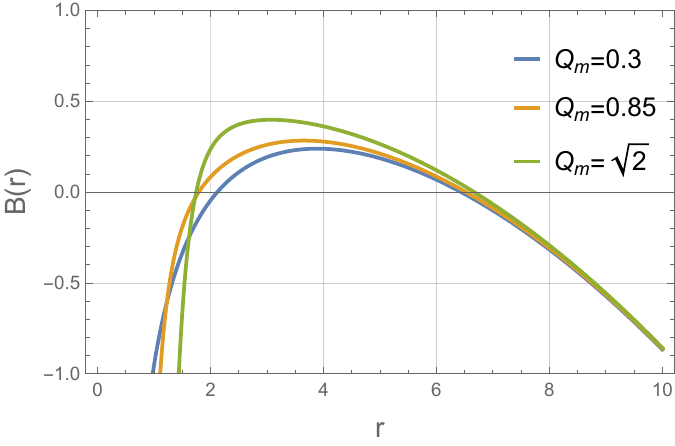}}
\caption{The metric function $B(r)$ as a function of the radial coordinate $r$ for different values of magnetic charge $Q_{\mathrm{m}}$ with fixed $M=1$ and $\Lambda=0.05$. Panels (a), (b), and (c) correspond to $\epsilon=-1$, $\epsilon=0$, and $\epsilon=1$, respectively. The zeros of $B(r)$ indicate horizons depending on parameters, there can be $0$--$3$ horizons (cosmological horizon $r_{\mathrm{c}}$, event horizon $r_{\mathrm{h}}$, and possibly Cauchy horizon $r_-$).}\label{fig31}
\end{figure}
\unskip
\begin{figure}[htb]
\centering
\subfigure[$Q_{\mathrm{m}}=0.3$]
{\label{B3} %% label for first subfigure
\includegraphics[width=2in]{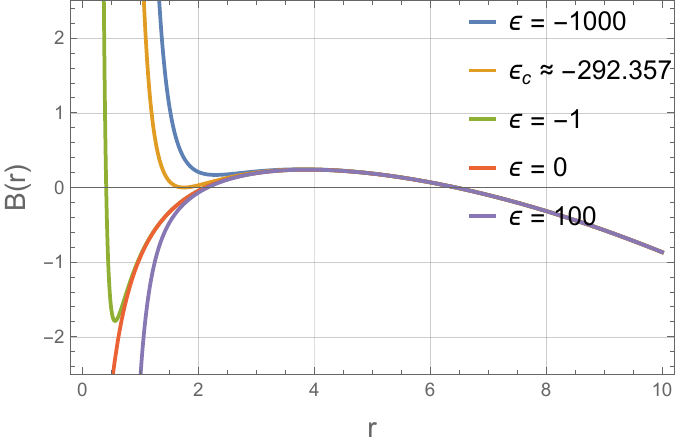}}
\hfill
\subfigure[$Q_{\mathrm{m}}=0.5$]
{\label{B6} %% label for first subfigure
\includegraphics[width=2in]{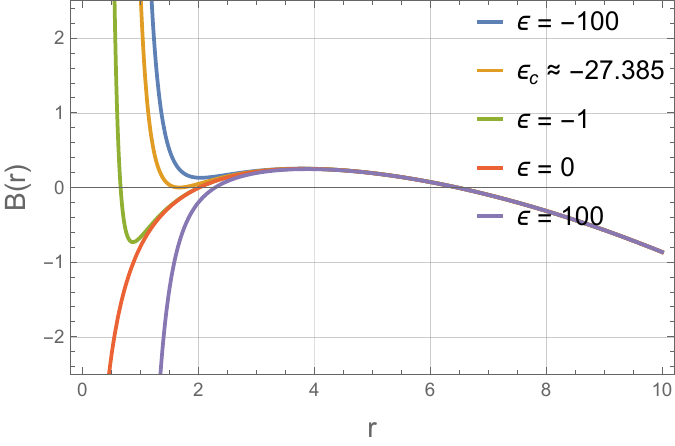}}
\hfill
\subfigure[$Q_{\mathrm{m}}=0.7 $]
{\label{B4} %% label for first subfigure
\includegraphics[width=2in]{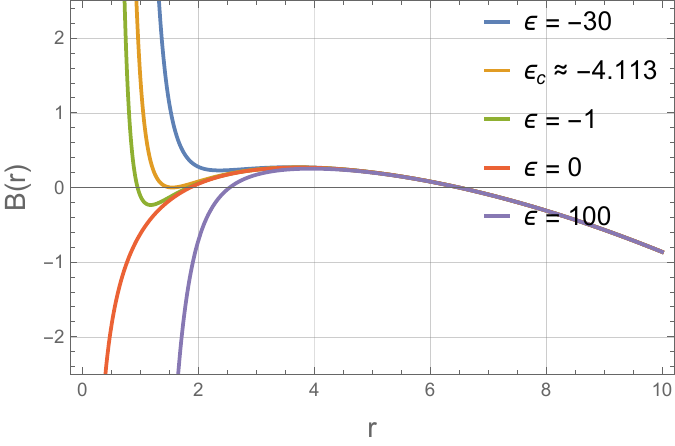}}
\caption{The metric function $B(r)$ versus $r$ for different values of the coupling parameter $\epsilon$,  with fixed $M=1$ and $\Lambda=0.05$.  Panels (a), (b), and (c) correspond to $Q_{\mathrm{m}}=0.3$, $Q_{\mathrm{m}}=0.5$, and $Q_{\mathrm{m}}=0.7$, respectively. The zeros of $B(r)$ indicate horizons depending on parameters, there can be $0$--$3$ horizons (cosmological horizon $r_{\mathrm{c}}$, event horizon $r_{\mathrm{h}}$, and possibly Cauchy horizon $r_-$).}\label{fig32}
\end{figure}
\unskip
\begin{figure}[htb]
\centering
\subfigure[$\epsilon=-1$ ]
{\label{B5} %% label for first subfigure
\includegraphics[width=2.1in]{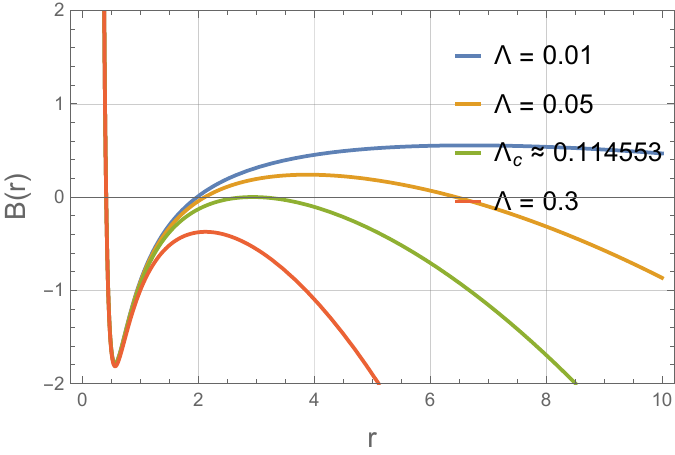}}
\hfill
\subfigure[$\epsilon=0$ ]
{\label{B7} %% label for first subfigure
\includegraphics[width=2.1in]{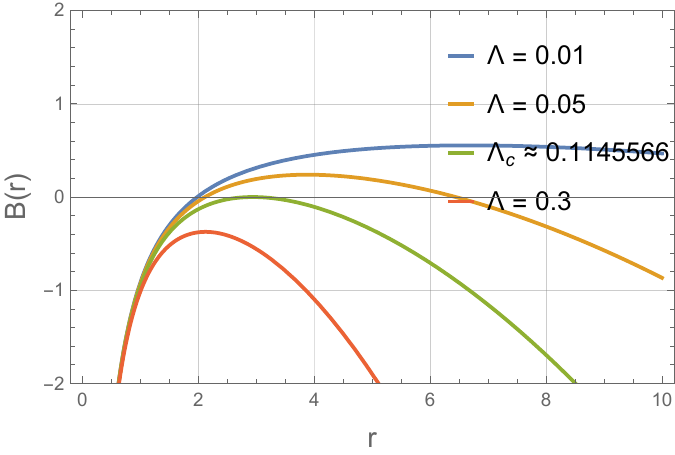}}
\hfill
\subfigure[$\epsilon=1$ ]
{\label{B8}%% label for first subfigure
\includegraphics[width=2.1in]{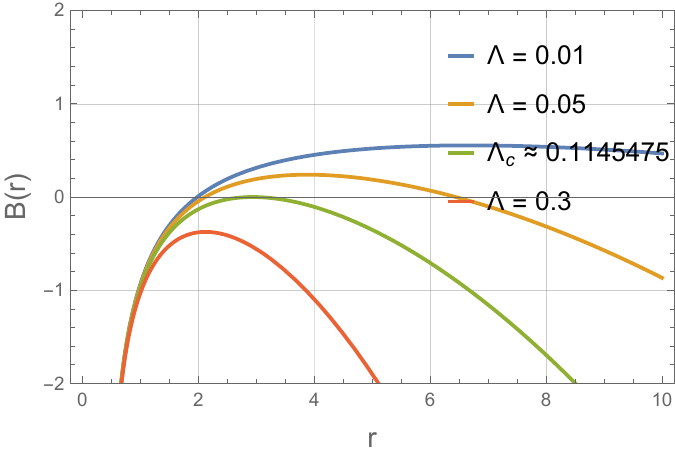}}
\caption{The metric function $B(r)$ versus $r$ for different values of the  cosmological constant $\Lambda$,  with fixed $M=1$ and  $Q_{\mathrm{m}}=0.3$. Panels (a), (b), and (c) correspond to $\epsilon=-1$, $\epsilon=0$, and $\epsilon=1$, respectively. The zeros of $B(r)$ indicate horizons depending on parameters, there can be $0$--$3$ horizons (cosmological horizon $r_{\mathrm{c}}$, event horizon $r_{\mathrm{h}}$, and possibly Cauchy horizon $r_-$).}\label{fig33}
\end{figure}

\section{Wave equations and perturbations}
\label{sec3}

In this section, we will show the master equations of various external fields, including the scalar field and electromagnetic field around magnetically charged de Sitter black hole. As noted in \cite{Konoplya:2011qq}, when the background remains unaltered by backreaction, black hole spacetime perturbations can be analyzed either by modifying the metric with perturbation terms or by introducing additional fields into the metric.

\subsection{Scalar field perturbation}

The equation of motion of a massless scalar field $\Phi$ in curved spacetime is given by the Klein-Gordon equation
\begin{eqnarray}   
\Box\Phi=\frac{1}{\sqrt{-g}}\partial_{\mu}\left(\sqrt{-g}g^{\mu\nu}\partial_{\nu}\Phi\right)= 0\label{eqscalar}.
\end{eqnarray}  

\begin{figure}[htb]
\centering
\subfigure[$\epsilon=-1$]
{\label{v11} %% label for first subfigure
\includegraphics[width=2in]{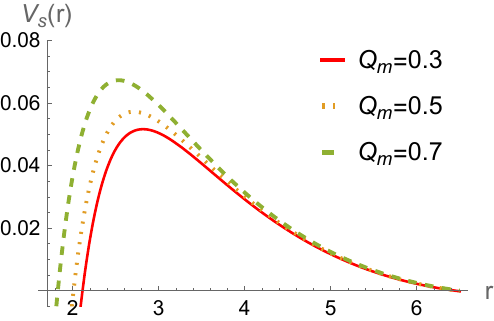}}
\hfill
\subfigure[$\epsilon=0$ ]
{\label{v12} %% label for first subfigure
\includegraphics[width=2in]{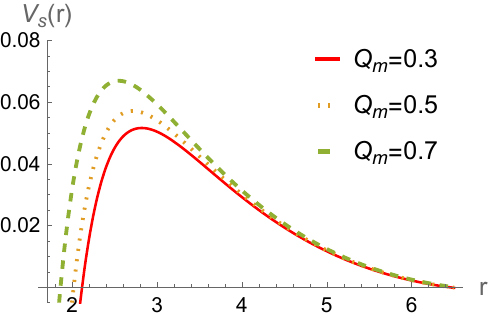}}
\hfill
\subfigure[ $\epsilon=1$ ]
{\label{v13} %% label for first subfigure
\includegraphics[width=2in]{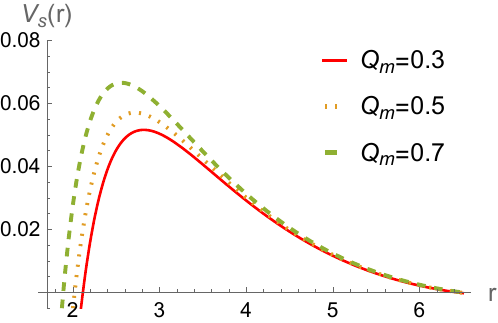}}
\hfill
\caption{The effective potential $V_{\mathrm{s}}(r)$ for massless scalar field perturbation ($l=1$) as a function of the radial coordinate $r$ for different values of magnetic charge $Q_{\mathrm{m}}$. In all cases, we set $M=1$ and $\Lambda=0.05$. Panels (a), (b), and (c) correspond to $\epsilon=-1$, $\epsilon=0$, and $\epsilon=1$, respectively. As $Q_{\mathrm{m}}$ decreases, the peak of the potential barrier shifts outward and its height is significantly suppressed.}\label{fig2}
\end{figure}

Substituting the metric \eq{metric} into the Klein-Gordon equation \eq{eqscalar}, and considering that the scalar field $\Phi(t, r, \theta, \varphi)$ propagates in a spherically symmetric background, we can separate the variables as follows:
\begin{eqnarray}  
\Phi(t, r, \theta, \varphi) =\sum_{l,m} \mathrm{e}^{-\mathrm{i} \omega t} \frac{\psi_{\mathrm{s}}(r)}{r}Y_{l,m}(\theta, \varphi) ,
 \end{eqnarray}   
where $Y_{l,m}(\theta, \varphi) $ are the spherical harmonics,  $\psi_{\mathrm{s}}(r)$ is radial wave function, $\omega$ is the frequency of $\Phi$, $l$ corresponds to the angular quantum number  of the black hole’s QNMs and then the perturbed field equation for the radial part in the tortoise coordinate can be written as
\begin{eqnarray} 
\frac{\mathrm{d}^2 \psi_{\mathrm{s}}(r_*)}{\mathrm{d}r_*^{2}} + \left[ \omega^2 - V_{\mathrm{s}}(r) \right] \psi_{\mathrm{s}}(r_*) = 0\label{E},  
\end{eqnarray} 
where the tortoise coordinate $r_*$ is defined as follows
\begin{eqnarray} 
\mathrm{d}r_*=\frac{1}{\sqrt{A(r)B(r)}}\mathrm{d}r, \label{rstar}
\end{eqnarray}
and $V_{\mathrm{s}}(r)$ is the effective potential 
\begin{eqnarray}  
V_{\mathrm{s}}(r)=\frac{ {A}'(r)B(r)+A(r){B}'(r)}{2r}+\frac{A(r)}{r^2}l(l+1). \label{potVs} 
\end{eqnarray}
%Evidently, this effective potential $V_{\mathrm{s}}(r)$ depends on the black hole background as well as angular quantum number $l$. 
Figs. \ref{fig2}, \ref{fig4}, and \ref{fig6}  illustrates that for scalar perturbations with a non-zero angular quantum number $(l > 0)$, an increase in the  magnetic charge $ Q_{\mathrm{m}}$ and the angular quantum number $ l $ results in a heightened potential barrier, while an increase in the coupling constant $\epsilon$ leads to a reduction in the potential's magnitude. However, in the distinct case of $ l = 0$, the effective potential profile exhibits a characteristic double-peak structure.

\subsection{Electromagnetic field perturbation}

Similar to Refs.\cite{Zhang:2025xqt,Yang:2023lcm}, we adopt the test field approximation for external electromagnetic perturbations, leading to the linear Maxwell equation in the curved background. The propagation of massless electromagnetic field in a curved spacetime, minimally coupled to the geometry, is driven by Maxwell’s equations
\begin{eqnarray}
  \frac{1}{\sqrt{-g}} \, \partial_\mu \left( \sqrt{-g} \, g^{\sigma\mu}g^{\rho\nu} F_{\rho\sigma}\right)=0, \label{eqmax}
\end{eqnarray}
where $F_{\rho\sigma}=\partial_{\rho}\boldsymbol{A}_{\sigma}-\partial_{\sigma}\boldsymbol{A}_{\rho}$  is the field strength tensor, and $\boldsymbol{A}_{\rho}$ is the vector potential  of the perturbed electromagnetic field which can be decomposed as
\begin{eqnarray}
  \boldsymbol{A}_\rho(t, r, \theta, \varphi) = \sum_{l,m} \mathrm{e}^{-\mathrm{i}\omega t}
\begin{bmatrix}
0 \\
0 \\
h_0(r)\frac{1} {\sin \theta} \frac{\partial Y_{l,m}}{\partial \varphi} \\
-h_0(r) \sin \theta \frac{\partial Y_{l,m}}{\partial \theta}
\end{bmatrix}
+ \sum_{l,m} \mathrm{e}^{-\mathrm{i}\omega t}
\begin{bmatrix}
h_1(r) Y_{l,m} \\
h_2(r) Y_{l,m} \\
h_3(r) \frac{\partial Y_{l,m}}{\partial \theta} \\
h_3(r) \frac{\partial Y_{l,m}}{\partial\varphi}
\end{bmatrix}. \label{eqA}
\end{eqnarray}
Here, $Y_{l,m}(\theta, \varphi) $ are spherical harmonics, and $l$ and $m$ are the angular and the azimuthal quantum numbers respectively. The first column in \eq{eqA} is the axial component with parity $(-1)^{l+1} $ and the second term is the polar mode with parity $(-1)^l$.

Substituting the axial and the polar modes of the electromagnetic perturbations into \eq{eqmax}, we can obtain 
\begin{eqnarray}
\frac{\mathrm{d}^2\Psi_{\mathrm{e}}(r_*)}{\mathrm{d}r_*^2} + \left(\omega^2 - V_{\mathrm{e}}(r)\right)\Psi_{\mathrm{e}}(r_*) = 0,\label{e}
\end{eqnarray}
where the potential is
\begin{eqnarray}
V_{\mathrm{e}}(r) = A(r) \frac{l(l+1)}{r^2} \label{potVe} ,
\end{eqnarray}
and the function \( \Psi_{\mathrm{e}}(r) \) takes different forms for the two modes
\begin{eqnarray}
&&\Psi_{\mathrm{e}}(r) = h_0(r), \quad \quad \quad  \textit{odd parity} ,\\
&&\Psi_{\mathrm{e}}(r) = -\sqrt{\frac{B(r)}{A(r)}}\frac{r^2}{l(l+1)} \left( \mathrm{i}\omega h_2(r) + \frac{\mathrm{d}h_1(r)}{\mathrm{d}r} \right) ,\quad \quad \textit{even parity} .
\end{eqnarray}

\begin{figure}[htb]
\centering
\subfigure[$\epsilon=-1$ ]
{\label{H11} %% label for first subfigure
\includegraphics[width=2in]{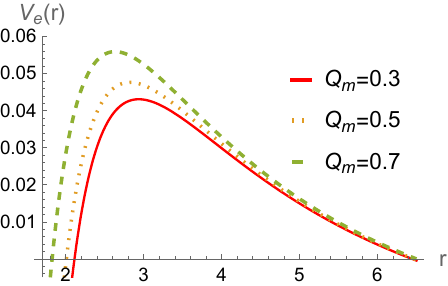}}
\hfill
\subfigure[$\epsilon=0$ ]
{\label{H21} %% label for first subfigure
\includegraphics[width=2in]{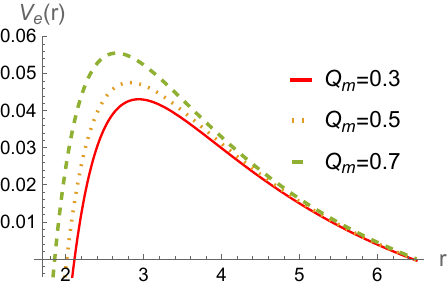}}
\hfill
\subfigure[ $\epsilon=1$ ]
{\label{H31} %% label for first subfigure
\includegraphics[width=2in]{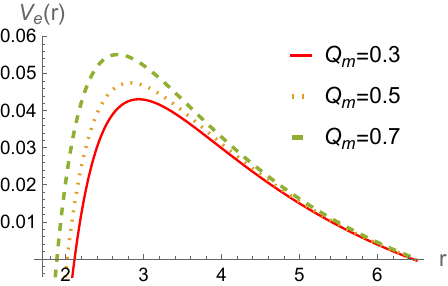}}
\caption{The effective potential $V_e(r)$ for electromagnetic field perturbation ($l=1$) as a function of the radial coordinate $r$ for different values of magnetic charge $Q_{\mathrm{m}}$. In all cases, we set $M=1$ and $\Lambda=0.05$. Panels (a), (b), and (c) correspond to $\epsilon=-1$, $\epsilon=0$, and $\epsilon=1$, respectively. As $Q_{\mathrm{m}}$ decreases, the peak of the potential barrier shifts outward and its height is significantly suppressed.}\label{fig3}
\end{figure}

\begin{figure}[H]
\centering
\subfigure[$\epsilon=-1$  ]
{\label{V21} %% label for first subfigure
\includegraphics[width=2in]{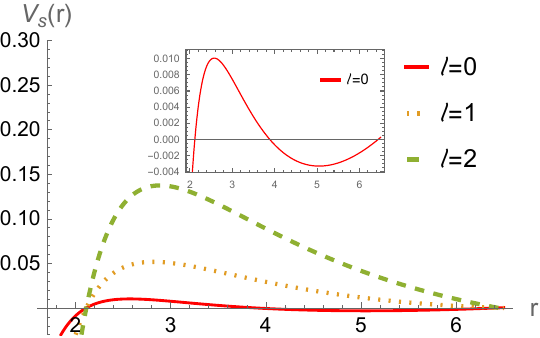}}
\hfill
\subfigure[$\epsilon=0$]
{\label{V22} %% label for first subfigure
\includegraphics[width=1.9in]{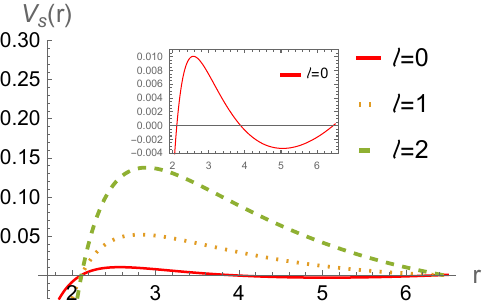}}
\hfill
\subfigure[ $\epsilon=1$]
{\label{V23} %% label for first subfigure
\includegraphics[width=2in]{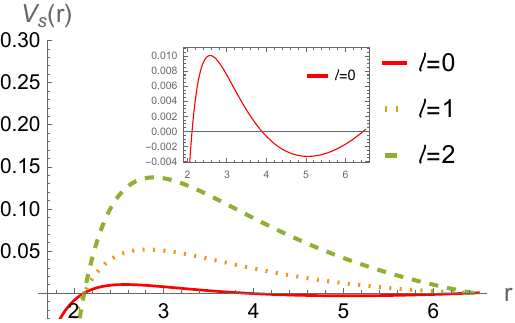}}
\hfill
\caption{The effective potential $V_{\mathrm{s}}(r)$ for massless scalar field perturbations on black hole with fixed $M=1$, $\Lambda=0.05$ and $Q_{\mathrm{m}}=0.3$. For these specific parameter values, the black hole event horizon is located at $r_{\mathrm{h}}\approx 2.11526$ and $r_{\mathrm{c}}\approx 6.44226$. Panels (a), (b), and (c) correspond to $\epsilon=-1$, $\epsilon=0$, and $\epsilon=1$, respectively. Each panel displays potentials for three values of the angular quantum number $l=0,1,2$.  Note the characteristic double-peak structure of the potential when $l=0$.}\label{fig4}
\end{figure}

\begin{figure}[H]
\centering
\subfigure[$\epsilon=-1$ ]
{\label{H12} %% label for first subfigure
\includegraphics[width=2in]{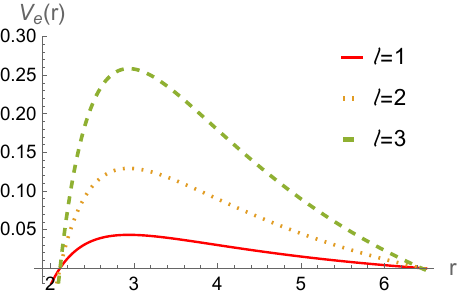}}
\hfill
\subfigure[$\epsilon=0$]
{\label{H22} %% label for first subfigure
\includegraphics[width=2in]{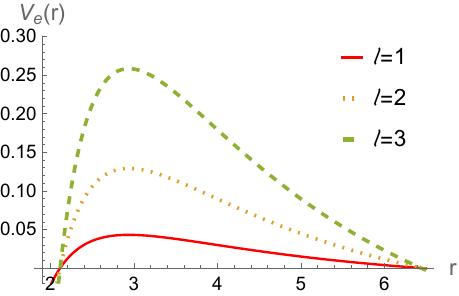}}
\hfill
\subfigure[ $\epsilon=1$]
{\label{H32} %% label for first subfigure
\includegraphics[width=2in]{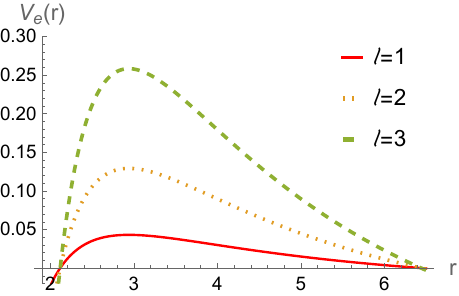}}
\caption{The effective potential $V_e(r)$ for electromagnetic field perturbation on black hole with fixed $M=1$, $\Lambda=0.05$ and $Q_{\mathrm{m}}=0.3$. Panels (a), (b), and (c) correspond to $\epsilon=-1$, $\epsilon=0$, and $\epsilon=1$, respectively. Each panel displays potentials for three values of the angular quantum number $l=1,2,3$. The potential barrier height increases monotonically with both the multipole number $l$ .}\label{fig5}
\end{figure}

\begin{figure}[H]
\centering
\subfigure[$Q_{\mathrm{m}}=0.3$]
{\label{Q12} %% label for first subfigure
\includegraphics[width=2in]{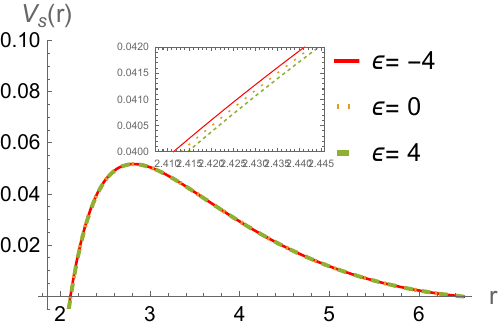}}
\hfill
\subfigure[$Q_{\mathrm{m}}=0.5$ ]
{\label{Q11} %% label for first subfigure
\includegraphics[width=2in]{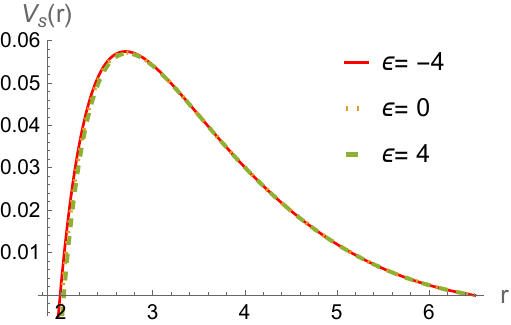}}
\hfill
\subfigure[$Q_{\mathrm{m}}=0.7$]
{\label{Q13} %% label for first subfigure
\includegraphics[width=2in]{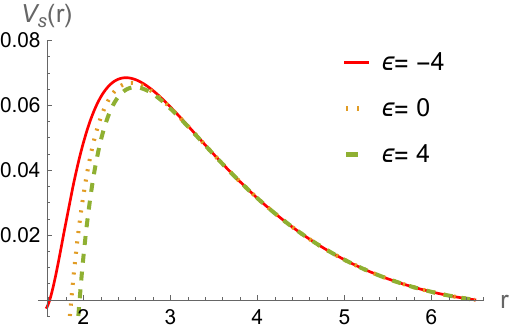}}
\caption{The effective potential $V_{\mathrm{s}}(r)$ for massless scalar field perturbation on black hole with fixed $M=1$,  $\Lambda=0.05$ and $l=1$. Panels (a), (b), and (c) correspond to $Q_{\mathrm{m}}=0.3$, $Q_{\mathrm{m}}=0.5$, and $Q_{\mathrm{m}}=0.7$, respectively. Each panel compares potentials for three coupling parameter values $\epsilon=-4$, $\epsilon=0$, and $\epsilon=4$. The potential barrier height decreases monotonically with increasing $\epsilon$.}\label{fig6}
\end{figure}

\begin{figure}[H]
\centering
\subfigure[$Q_{\mathrm{m}}=0.3$]
{\label{Q21} %% label for first subfigure
\includegraphics[width=2in]{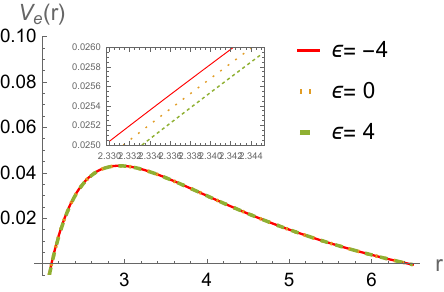}}
\hfill
\subfigure[$Q_{\mathrm{m}}=0.5$ ]
{\label{Q22} %% label for first subfigure
\includegraphics[width=2in]{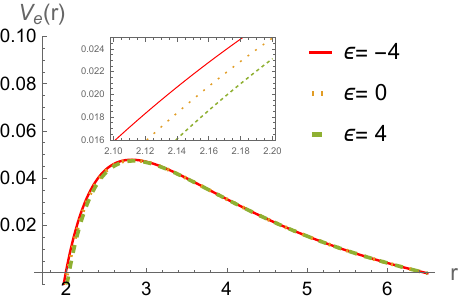}}
\hfill
\subfigure[$Q_{\mathrm{m}}=0.7$]
{\label{Q23} %% label for first subfigure
\includegraphics[width=2in]{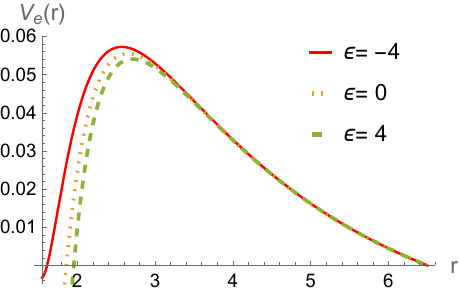}}
\caption{The effective potential $V_e(r)$ for electromagnetic field perturbation on black hole with fixed $M=1$,  $\Lambda=0.05$ and $l=1$. Panels (a), (b), and (c) correspond to $Q_{\mathrm{m}}=0.3$, $Q_{\mathrm{m}}=0.5$, and $Q_{\mathrm{m}}=0.7$, respectively. Each panel compares potentials for three coupling parameter values $\epsilon=-4$, $\epsilon=0$, and $\epsilon=4$. The potential barrier height decreases monotonically with increasing $\epsilon$.}\label{fig60}
\end{figure}

The radial profile of the effective potential $V_e(r)$ exhibits distinct characteristics for different combinations of the  magnetic charge $Q_{\mathrm{m}}$, coupling constant $\epsilon$, and angular quantum number $l$, as depicted in Figs.\ref{fig3}, \ref{fig5}, and \ref{fig60}. Crucially, the potential maintains positive definiteness throughout the exterior spacetime region ($ r_{\mathrm{h}}<r <r_{\mathrm{c}}$) for all parameter sets considered. This persistent positivity, particularly the absence of any negative potential well, provides strong evidence for the black hole's structural stability against massless scalar field perturbations. Furthermore, quantitative analysis reveals that the potential barrier's height and width demonstrate monotonic enhancement with increasing values of $ Q_{\mathrm{m}}$, $\epsilon$, and $l$, suggesting a parametric dependence of the scattering properties on these fundamental parameters.

It is important to note that for electromagnetic perturbations, the angular quantum number $l$ starts from $l=1$. This is because the monopole mode ($l=0$) corresponds to a static Coulomb field which is non-dynamical and does not emit radiation. Therefore, the lowest radiative mode for the electromagnetic field is the dipole mode ($l=1$).



\section{Quasinormal Mode frequencies}
\label{sec4}

To accurately determine the QNMs frequencies of magnetically charged black holes, we employ two distinct computational approaches: AIM and  WKB approximation method. This dual-methodology approach enables us to perform cross-validation of our results.

\subsection{Asymptotic iteration method}

In this section, we will show the main steps of the AIM method. Firstly, we rewrite scalar perturbed equation \eqref{E} in terms of $u=1/r$
\begin{eqnarray}
&&\frac{1}{A(u)B(u)}\Big[\frac{1}{u}\left(\frac{2\omega^2}{u}+u^2B(u)A'(u)+u^2A(u)B'(u)\right)-2l(l+1)\Big]\psi(u)\nonumber\\
&&\Big[4u+\frac{u^2(B(u)A'(u)-A(u)B'(u))}{A(u)B(u)}\Big]\psi'(u)+2u^2\psi''(u).\label{equ1}
\end{eqnarray}
To factor out the divergent behavior at both the event horizon and the cosmological horizon, we introduce the following transformation to ensure the regularity of the wave function:
\begin{eqnarray}
  \psi(u)=\mathrm{e}^{\mathrm{i}\omega r_*} \chi(u)\label{eqschi}.
\end{eqnarray}
where $\chi(u)$ should be a finite and convergent function, to establish the proper scaling behavior for quasinormal mode boundary conditions, we define
\begin{eqnarray}
\mathrm{e}^{\mathrm{i}\omega r_*}=\prod_{j}(u-u_j)^{\mathrm{i}\omega/\kappa_j},
\end{eqnarray}
where $\kappa_j$ denotes the surface gravity evaluated at the coordinate $u_j$, defined by the condition $B(u)|_{u=u_j}=0$ . This scaling transformation regularizes the solution by removing the divergent behavior near the $u_j$ boundary and simultaneously enforces the boundary conditions. We introduce the surface gravity at the event horizon
\begin{eqnarray}
\kappa_{\mathrm{h}}=\frac{1}{2}\frac{ \mathrm{d}\sqrt{A(r)B(r)}}{\mathrm{d} r}\mid_{r=r_{\mathrm{h}}}=\frac{1}{2}(\sqrt{A(r_{\mathrm{h}})B(r_{\mathrm{h}})})'.
\end{eqnarray}
Substitute \eq{eqschi} to \eq{equ1}, we have
\begin{eqnarray}
  \chi''=\lambda_0(u)\chi'+s_0(u)\chi \label{eqschieq},
\end{eqnarray}
where 
\begin{eqnarray}
  \lambda_0(u)=\frac{1}{2}\Big[-\frac{4}{u}-\frac{4\mathrm{i} r_{\mathrm{c}}\omega}{\kappa_{\mathrm{c}}-\kappa_{\mathrm{c}} r_{\mathrm{c}} u}-\frac{4\mathrm{i} r_{\mathrm{h}} \omega}{\kappa_{\mathrm{h}}-\kappa_{\mathrm{h}} r_{\mathrm{h}} u}-\frac{A'(u)}{A(u)}-\frac{B'(u)}{B(u)}\Big],
\end{eqnarray}
and
\begin{eqnarray}
  s_0(u)&=&\frac{1}{2\kappa_{\mathrm{c}}^2\kappa_{\mathrm{h}}^2u^4(r_{\mathrm{c}} u-1)^2(r_{\mathrm{h}} u-1)^2A(u)B(u)}\Big[-\kappa_{\mathrm{c}}\kappa_{\mathrm{h}}(r_{\mathrm{c}} u-1)(r_{\mathrm{h}} u-1) \nonumber\\
  &&\times\Big(2\kappa_{\mathrm{c}} \kappa_{\mathrm{h}}(r_{\mathrm{c}} u-1)(r_{\mathrm{h}} u-1)\omega^2+u^3\Big(\kappa_{\mathrm{c}} \kappa_{\mathrm{h}}(r_{\mathrm{c}} u-1)(r_{\mathrm{h}} u-1)\nonumber\\
  &&-\mathrm{i} u( \kappa_{\mathrm{h}} r_{\mathrm{c}} - \kappa_{\mathrm{c}} r_{\mathrm{h}}+(\kappa_{\mathrm{c}}+\kappa_{\mathrm{h}})r_{\mathrm{c}} r_{\mathrm{h}} u)\omega\Big)B(u)A'(u)\Big)\nonumber\\
  &&+u^2 A(u)\Big(2 u \omega \Big(-\mathrm{i} \kappa_{\mathrm{c}}\kappa_{\mathrm{h}}(-\mathrm{i}\kappa_{\mathrm{c}} \kappa_{\mathrm{h}}(r_{\mathrm{c}} u-1)^2(r_{\mathrm{h}} u-2)-\kappa_{\mathrm{h}}\kappa_{\mathrm{c}}(r_{\mathrm{c}} u-2)(r_{\mathrm{h}} u-1)^2)\nonumber\\
  &&+u(\kappa_{\mathrm{h}} r_{\mathrm{c}}+\kappa_{\mathrm{c}} r_{\mathrm{h}}-(\kappa_{\mathrm{c}}+\kappa_{\mathrm{h}})r_{\mathrm{c}} r_{\mathrm{h}} u)^2\omega\Big)B(u)+\kappa_{\mathrm{c}}\kappa_{\mathrm{h}}(r_{\mathrm{c}} u-1)(r_{\mathrm{h}} u-1)\nonumber\\
  &&\times\big(2\kappa_{\mathrm{c}}\kappa_{\mathrm{c}} l(l+1)(r_{\mathrm{c}} u-1)(r_{\mathrm{h}} u-1)+u(-\kappa_{\mathrm{c}}\kappa_{\mathrm{h}}(r_{\mathrm{c}} u-1)(r_{\mathrm{h}} u-1)\nonumber\\
  &&\mathrm{i} u (-\kappa_{\mathrm{h}} r_{\mathrm{c}}-\kappa_{\mathrm{c}} r_{\mathrm{h}}+(\kappa_{\mathrm{c}}+\kappa_{\mathrm{h}})r_{\mathrm{c}} r_{\mathrm{h}} u)\omega)B'(u)\Big)\Big)\Big].
\end{eqnarray}
Based on $\lambda_0$ and $s_0$, the perturbed equation \eqref{eqschieq} can be solved numerically by using the improved AIM \cite{Cho:2009cj}. 
Following the same procedure described above, we can also obtain functions $\lambda_0$ and $s_0$ for the
electromagnetic field perturbation:
\begin{eqnarray}
  \lambda_0(u)=\frac{1}{2}\Big[-\frac{4}{u}-\frac{4\mathrm{i} r_{\mathrm{c}}\omega}{\kappa_{\mathrm{c}}-\kappa_{\mathrm{c}} r_{\mathrm{c}} u}-\frac{4\mathrm{i} r_{\mathrm{h}} \omega}{\kappa_{\mathrm{h}}-\kappa_{\mathrm{h}} r_{\mathrm{h}} u}-\frac{A'(u)}{A(u)}-\frac{B'(u)}{B(u)}\Big],
\end{eqnarray}
and
\begin{eqnarray}
  s_0(u)&=&\frac{1}{2\kappa_{\mathrm{c}}^2\kappa_{\mathrm{h}}^2u^4(r_{\mathrm{c}} u-1)^2(r_{\mathrm{h}} u-1)^2A(u)B(u)}\Big[\kappa_{\mathrm{c}}\kappa_{\mathrm{h}} \omega(r_{\mathrm{c}} u-1)(r_{\mathrm{h}} u-1)\nonumber\\
  &&\times\Big(-2\kappa_{\mathrm{c}} \kappa_{\mathrm{h}}(r_{\mathrm{c}} u-1)(r_{\mathrm{h}} u-1)\omega+\mathrm{i} u^4\Big(- \kappa_{\mathrm{h}} r_{\mathrm{c}} - \kappa_{\mathrm{c}} r_{\mathrm{h}}+(\kappa_{\mathrm{c}}+\kappa_{\mathrm{h}})r_{\mathrm{c}} r_{\mathrm{h}} u)\omega\Big)B(u)A'(u)\Big)\nonumber\\
  &&+u^2 A(u)\Big(2 u \omega \Big(-\mathrm{i} \kappa_{\mathrm{c}}\kappa_{\mathrm{h}}(-\mathrm{i}\kappa_{\mathrm{c}} \kappa_{\mathrm{h}}(r_{\mathrm{c}} u-1)^2(r_{\mathrm{h}} u-2)-\kappa_{\mathrm{h}}\kappa_{\mathrm{c}}(r_{\mathrm{c}} u-2)(r_{\mathrm{h}} u-1)^2)\nonumber\\
  &&+u(\kappa_{\mathrm{h}} r_{\mathrm{c}}+\kappa_{\mathrm{c}} r_{\mathrm{h}}-(\kappa_{\mathrm{c}}+\kappa_{\mathrm{h}})r_{\mathrm{c}} r_{\mathrm{h}} u)^2\omega\Big)B(u)+\kappa_{\mathrm{c}}\kappa_{\mathrm{h}}(r_{\mathrm{c}} u-1)(r_{\mathrm{h}} u-1)\nonumber\\
  &&\times \Big(2\kappa_{\mathrm{c}}\kappa_{\mathrm{c}} l(l+1)(r_{\mathrm{c}} u-1)(r_{\mathrm{h}} u-1)+\mathrm{i} u^2 (-\kappa_{\mathrm{h}} r_{\mathrm{c}}-\kappa_{\mathrm{c}} r_{\mathrm{h}}+(\kappa_{\mathrm{c}}+\kappa_{\mathrm{h}})r_{\mathrm{c}} r_{\mathrm{h}} u)\omega)B'(u)\Big)\Big)\Big].\nonumber\\
\end{eqnarray}

\subsection{6th-order WKB method}

This method, first proposed by Schutz and Will, was used to address black hole scattering problems \cite{Kokkotas:1988fm}. Later, further developments were made by Iyer, Will, and Konoplya  \cite{Konoplya:2011qq}. In this paper, we consider the most commonly used 6th-order WKB approximation method \cite{Konoplya:2011qq}
\begin{align}
\frac{\mathrm{i}(\omega^2 - V_0)}{\sqrt{-2V_0''}} - \sum_{i=2}^6 \Lambda_i = n + \frac{1}{2}, \quad (n = 0, 1, 2, \cdots)
\end{align}
where \( V''(r_0) \) is the value of the second derivative of the effective potential with respect to \( r \) at its maximum point \( r_0 \) defined by the solution of the equation \( \left. \frac{\mathrm{d}V}{\mathrm{d}r_*} \right|_{r_*=r_0}= 0 \). \( V_0 \) represents the maximum value of the effective potential, and \( \Lambda_i \) is the \( i \)-th order revision terms depending on the values of the effective potential. This semi-analytical method has been applied extensively in numerous black hole spacetime cases. 
It should be pointed out here the WKB approach works well for situations where the multipole number is larger compared to the overtone: $l\geq n$, the WKB approach produces less accurate outcomes for $l<n$ \cite{Iyer:1986nq,Konoplya:2003ii}. 

\subsection{Bernstein spectral method}

In light of the limitations of the WKB method in handling the double-peaked potential observed for the $l=0$ scalar perturbations, we employ the high-precision Bernstein spectral method. This numerical approach converts the differential boundary value problem into a matrix eigenvalue problem by expanding the solution in a basis of Bernstein polynomials. We first introduce a compact coordinate $v$ that maps the physical region between the black hole's event horizon $r_{\mathrm{h}}$, and the cosmological horizon $r_{\mathrm{c}}$\cite{Fortuna:2020obg}, onto the unit interval $[0,1]$:
\begin{eqnarray}
v \equiv \frac{1/r - 1/r_{\mathrm{c}}}{1/r_{\mathrm{h}} - 1/r_{\mathrm{c}}}.
\end{eqnarray}
In this coordinate system, the cosmological horizon $r_{\mathrm{c}}$ corresponds to $v=0$, and the event horizon $r_{\mathrm{h}}$ corresponds to $v=1$.

The wave function is factorized as
$\Psi(v) = F(v,\omega)\psi(v)$, with $F(v,\omega)$ encoding the near-horizon asymptotics such that $\psi(v)$ is regular at both $v=0$ and $v=1$ when $\omega$ is a quasinormal frequency. We then approximate this regular function $\psi(v)$ as a finite sum of $N$-th degree Bernstein polynomials:
\begin{eqnarray}
\psi(v) \approx \sum_{k=0}^N c_k, B_{k,N}(v), \qquad
B_{k,N}(v) = \binom{N}{k} v^k (1-v)^{N-k}.
\end{eqnarray}
This expansion is substituted into the radial wave equation, which is then evaluated on a grid of $N+1$ collocation points. The Chebyshev collocation grid\cite{Dias:2010eu,Jansen:2017oag} is typically chosen for its excellent convergence properties:
\begin{eqnarray}
v_p = \sin^2\left(\frac{p\pi}{2N}\right), \qquad p=0,1,\dots,N.
\end{eqnarray}
This procedure transforms the differential equation into a set of linear algebraic equations for the coefficients $c_k$. This system has a non-trivial solution if and only if its coefficient matrix, whose elements are polynomials of the frequency $\omega$, is singular. Thus, the problem of finding the quasinormal mode frequencies $\omega$ is reduced to a generalized matrix eigenvalue problem, which can be solved efficiently using numerical methods.

The use of a finite polynomial basis can introduce non-physical, "spurious" eigenvalues that do not converge as the basis size increases. To ensure the reliability of our results, we check for the convergence of the solutions by systematically increasing the polynomial degree $N$. Only the eigenvalues that remain stable and agree within a desired precision across different values of $N$ are identified as the true quasinormal mode frequencies\cite{Konoplya:2022xid}.

\section{Numerical results}
\label{sec5}

For the scalar perturbation with $l=0$, the effective potential may develop a double-peak structure, where the standard WKB approximation—based on expansion around a single maximum—becomes unreliable. To obtain accurate quasinormal frequencies in this sector, we apply the Bernstein spectral method. We find that the Bernstein results agree well with AIM values, while the WKB approximation shows significant deviations.
We also calculate the percentage deviation $\varepsilon_1$ (or $\varepsilon_2$)
of QNFs obtained via the AIM and Bernstein spectral method (or WKB methods). The relative error between the two methods is defined by
\begin{eqnarray}
\varepsilon_1=\frac{|\omega_\mathrm{AIM}-\omega_\mathrm{BP}|}{|\omega_\mathrm{AIM}|}\times 100\%,~~\varepsilon_2=\frac{|\omega_\mathrm{AIM}-\omega_\mathrm{WKB}|}{|\omega_\mathrm{AIM}|}\times 100\%.
\end{eqnarray}
To ensure the accuracy and convergence of the calculated QNFs, we employed $N=30$ and $N=40$ iterations of the Bernstein spectral method for cross-validation when determining the true values for the $l=0$ scalar field.

\subsection{$Q_{\mathrm{m}}$-dependence }

Table \ref{S0} and Table \ref{E1} present the fundamental quasinormal modes (QNFs) for scalar and electromagnetic perturbations ($n=0$) with fixed black hole mass $M=1$ and cosmological constant $\Lambda=0.05$. We investigate the cases for angular quantum numbers $l=0, 1$ in scalar perturbation and $l=1$ in electromagnetic perturbation, analyzing the influence of the coupling parameter $\epsilon$ and the magnetic charge $Q_{\mathrm{m}}$.

\begin{figure}[H]
\centering
\subfigure[$\epsilon=1$]
{ \label{s3r}%% label for first subfigure
\includegraphics[width=2.5in]{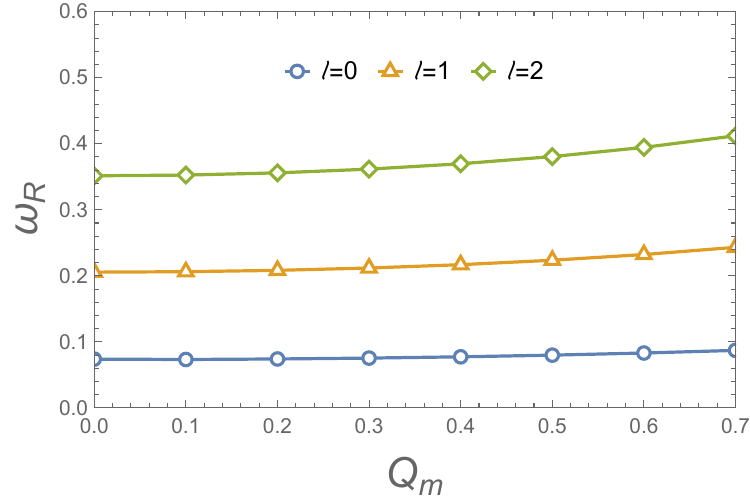}}
\subfigure[$\epsilon=1$]
{\label{s3i} %% label for first subfigure
\includegraphics[width=2.55in]{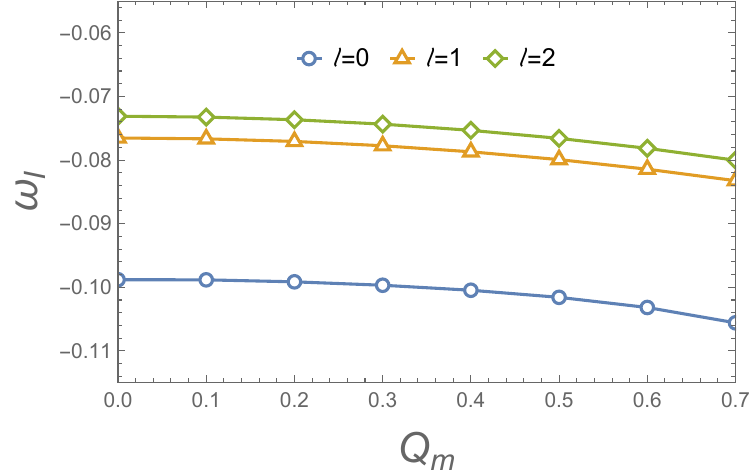}}
\subfigure[$\epsilon=0$]
{ \label{s2r}%% label for first subfigure
\includegraphics[width=2.5in]{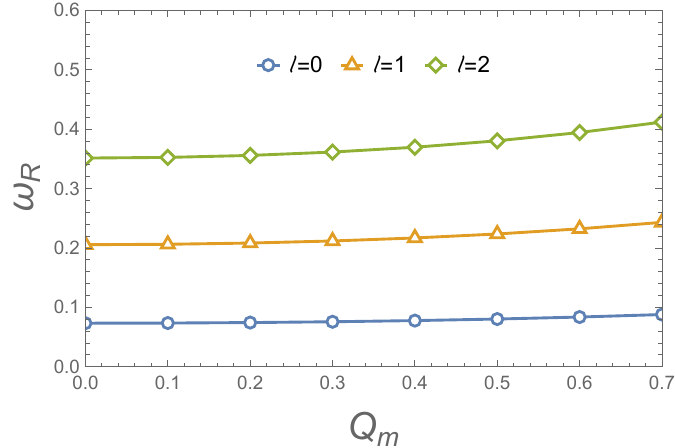}}
\subfigure[$\epsilon=0$]
{\label{s2i} %% label for first subfigure
\includegraphics[width=2.55in]{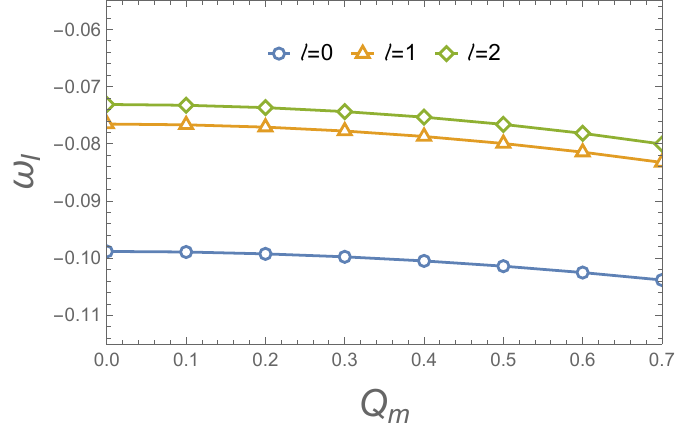}}
\subfigure[$\epsilon=-1$]
{ \label{s1r}%% label for first subfigure
\includegraphics[width=2.5in]{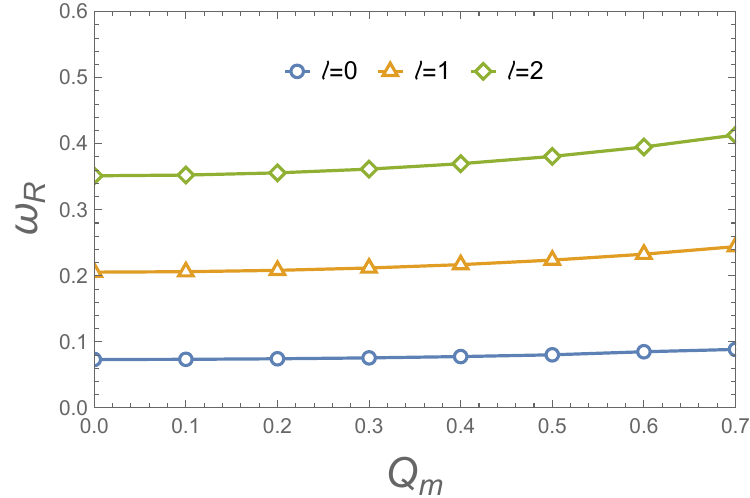}}
\subfigure[$\epsilon=-1$]
{\label{s1i} %% label for first subfigure
\includegraphics[width=2.55in]{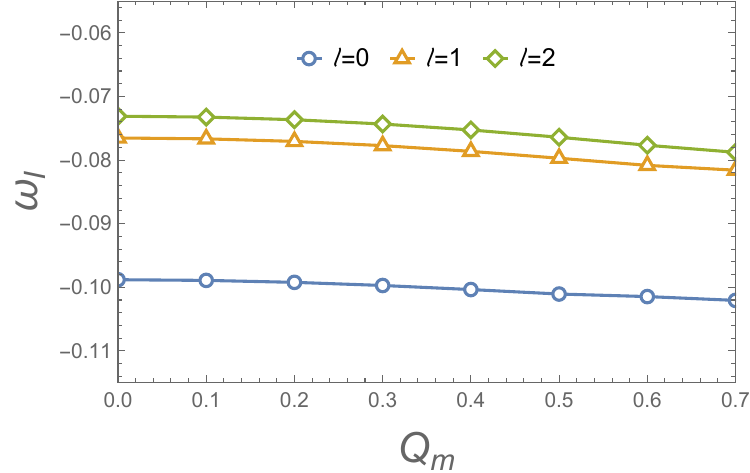}}
\caption{Variation of scalar fundamental QNFs ($n=0$) with respect to the magnetic charge $Q_{\mathrm{m}}$ for black hole with fixed mass $M=1$ and cosmological constant $\Lambda=0.05$. Panels (a,b) correspond to coupling parameter $\epsilon=1$, (c,d) to $\epsilon=0$, and (e,f) to $\epsilon=-1$. In each panel, curves for angular quantum numbers $l=0,1,2$ are shown. For all coupling parameters considered, both $\text{Re}(\omega)$ and $|\text{Im}(\omega)|$ increase monotonically with growing magnetic charge $Q_{\mathrm{m}}$.}\label{fig11}
\end{figure}

\begin{figure}[H]
\centering
\subfigure[$\epsilon=-1$]
{ \label{e3r}%% label for first subfigure
\includegraphics[width=2.5in]{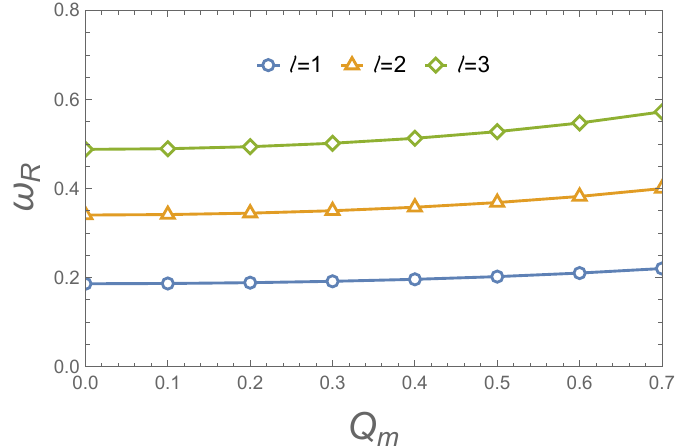}}
\subfigure[$\epsilon=-1$]
{\label{e3i} %% label for first subfigure
\includegraphics[width=2.55in]{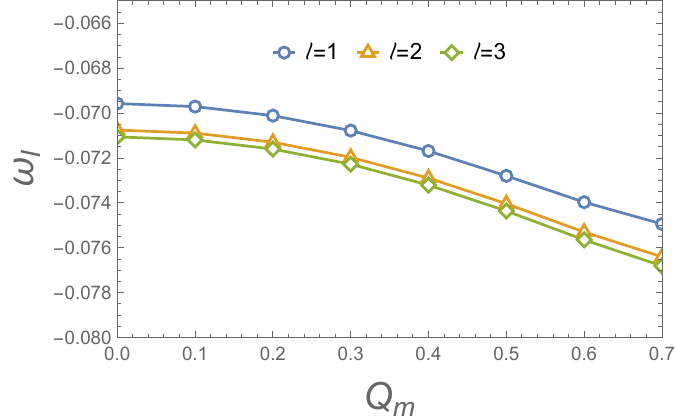}}
\subfigure[$\epsilon=0$]
{ \label{e2r}%% label for first subfigure
\includegraphics[width=2.5in]{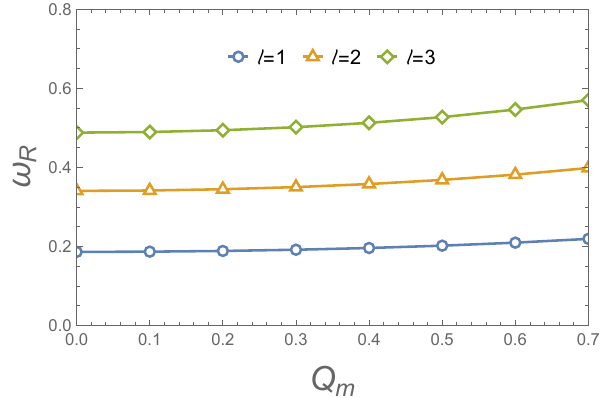}}
\subfigure[$\epsilon=0$]
{\label{e2i} %% label for first subfigure
\includegraphics[width=2.55in]{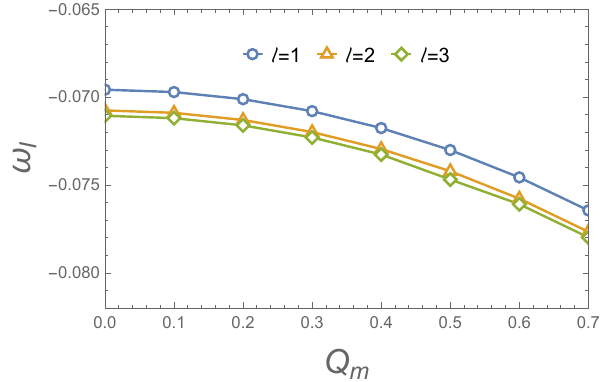}}
\subfigure[$\epsilon=1$]
{ \label{e1r}%% label for first subfigure
\includegraphics[width=2.5in]{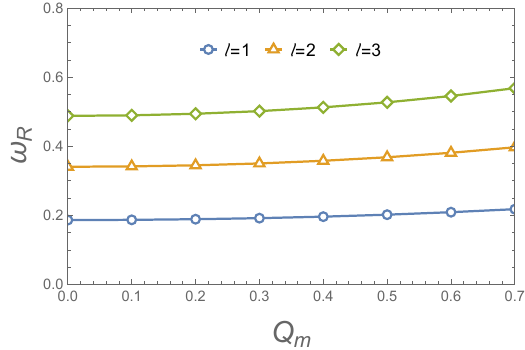}}
\subfigure[$\epsilon=1$]
{\label{e1i} %% label for first subfigure
\includegraphics[width=2.55in]{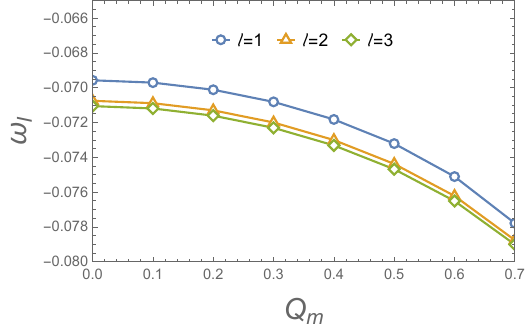}}
\caption{Variation of electromagnetic fundamental QNFs ($n=0$) with respect to the magnetic charge $Q_{\mathrm{m}}$ for black hole with fixed mass $M=1$ and cosmological constant $\Lambda=0.05$. Panels (a,b) correspond to coupling parameter $\epsilon=-1$, (c,d) to $\epsilon=0$, and (e,f) to $\epsilon=1$. In each panel, curves for angular quantum numbers $l=1,2,3$ are shown. For all coupling parameters considered, both $\text{Re}(\omega)$ and $|\text{Im}(\omega)|$ increase monotonically with growing magnetic charge $Q_{\mathrm{m}}$.}\label{fig22}
\end{figure}

Figs.\ref{fig11} illustrates the behavior of the real part (oscillation frequency) and the imaginary part (damping rate) of the fundamental QNFs ($n=0$) for scalar perturbations as a function of the magnetic charge $Q_{\mathrm{m}}$. The results for $l=0,1,2$ are shown. As can be seen from the Figs.\ref{s3r}\ref{s2r}\ref{s1r}, for all considered values of the coupling parameter $\epsilon=0,\pm 1$, the real part of the QNFs, $\text{Re}(\omega)$ monotonically increases with an increase in $Q_{\mathrm{m}}$. This implies that a larger magnetic charge leads to a higher oscillation frequency for the scalar field perturbation. The behavior of the imaginary part, $|\text{Im}(\omega)|$, increases significantly with $Q_{\mathrm{m}}$, indicating that the perturbations decay faster, as shown in Figs.\ref{s3i}\ref{s2i}\ref{s1i}. Figs.\ref{fig22} show similar trends for electromagnetic perturbations, with results presented for $l=1,2,3$. The behavior of the electromagnetic QNFs is qualitatively similar to that of the scalar case. These results demonstrate that the black hole's magnetic charge $Q_{\mathrm{m}}$, is a crucial physical parameter influencing its quasinormal mode spectrum. It not only alters the oscillation frequency of perturbations but also affect their decay time.

\begin{table}[h!]
\caption{Comparison of fundamental QNFs $(n=0)$ for massless scalar field perturbations $l=0$ obtained via the AIM, Bernstein spectral and WKB method. The parameters are fixed as $M=1$, $\Lambda=0.05$, $\epsilon=0, \pm 1$ and $Q_{\mathrm{m}}$ varies from $0.3$ to $0.7$. The parameter $\varepsilon_1$ represents relative errors between AIM and Bernstein methods, $\varepsilon_2$ represents relative errors between AIM and WKB methods.}\label{S0}
\resizebox{\linewidth}{!}{
\begin{tabular}{|c|c|c|c|c|c|c|} \hline
$\epsilon$ &  	$Q_{\mathrm{m}}$ & AIM  & Bernstein  spectral  & $\varepsilon_1(\%)$  &    WKB  & $\varepsilon_2(\%)$ \\ \hline
 \multirow{3}{*}{1}& 0.3 &  $0.0755244 - 0.0996825 i $&$0.0754839 - 0.0996719  i$   & 0.033532     &   $0.0703929 - 0.0992877 i    $  &  4.1153 \\  
 & 0.5 &  $0.0799900 -0.1015850 i$ &$0.0799338 -0.1015621  i$   &  0.046942 & $0.0760739 - 0.0984478 i$ &3.8808  \\ 
 &  0.7 &  $0.0872094 - 0.1055948 i$ &$0.0870846 - 0.1055414i $   &  0.099110 &  $0.0927763 - 0.0944930 i$ & 9.0684 \\ \hline
 \multirow{3}{*}{0}& 0.3 &  $0.0761242 - 0.0999622 i$  & $0.0755935 - 0.099640      i$    &0.49392    & $0.0711911 - 0.0982916 i$  &4.1452 \\ 
&  0.5 &  $0.0803893 - 0.101386 i$ &$0.0800 - 0.1013  i$   &  0.46392 & $0.0762451 - 0.0995951  i$  & 3.4893 \\ 
&  0.7 &  $0.0874256 - 0.103788  i$ &$0.0870 - 0.1040    i$   &  0.35025 &$0.0840822 - 0.10198 i$  &2.8011 \\ \hline
\multirow{3}{*}{-1}& 0.3 &  $0.0755758 - 0.0996304 i$  & $0.0755 - 0.0996    i$    &0.076028 & $0.0697726 - 0.100461 i$   &4.6880    \\ 
&  0.5 &  $0.0800608 - 0.100936  i$ &$0.0800631 - 0.101291   i$   & 0.27597 &  $0.0768723 - 0.0999537  i$ & 2.5896 \\ 
&  0.7 &  $0.0874665 - 0.101225 i$ &$0.0874479 - 0.101187 i$   & 0.031672&   $0.0883847 - 0.0975922 i$ & 2.8012\\ \hline
\end{tabular}}
\end{table}

\begin{table}[h!]
\caption{The fundamental QNFs ($n=0$) of scalar perturbations and electromagnetic perturbations $l=1$ 
for various $\epsilon$ and $Q_{\mathrm{m}}$. Other parameters are set to $M=1$ and $\Lambda=0.05$. The parameter $\varepsilon_2$ represents relative errors between AIM and WKB methods.}\label{E1}
\resizebox{\linewidth}{!}{
\begin{tabular}{|c|c|c|c|c|c|c|c|} \hline
&  & \multicolumn{3}{|c|}{scalar perturbation} & \multicolumn{3}{|c|}{electromagnetic perturbation}   \\ \hline
$\epsilon$ &  	$Q_{\mathrm{m}}$ & AIM  &  WKB  & $\varepsilon_2(\%)$ & AIM  &  WKB  & $\varepsilon_2(\%)$ \\ \hline
 \multirow{3}{*}{1} & 0.3 &  
 $0.211816 - 0.0777350i $&$0.211845- 0.0777674i$    &   0.019117 &
 $0.191863-0.0707426i $&$0.191835- 0.0708107i$    &    0.036162 \\ 
  & 0.5 &  
  $0.223456 - 0.0801152i$ &$0.0845594 - 0.109569i$   &  0.011044  &
 $  0.201987-0.0731151i$ &$0.201913 -0.0732115i$   &  0.056637 \\
   & 0.7 &  
 $0.242213 - 0.0845578i $&$ 0.24207 -0.084608i$    &0.058903  &
 $  0.218185-0.0775617i $&$ 0.217855- 0.0777876i$    &0.17296 \\  \hline
  \multirow{3}{*}{0}
 & 0.3 &  
 $0.2118342-0.0777156 i$&$0.2118612 - 0.0777493 i$    & 0.019140   &
 $0.1918838-0.0707257 i$&$0.191852 -0.0707951 i$    & 0.037424\\ 
  & 0.5 &  
  $0.2236277-0.0799076i$ &$ 0.2236487 - 0.0799311i$   &   0.013271 &
  $0.2021921-0.0729303i$ &$ 0.202161 - 0.0730071i$   &   0.038590\\
   & 0.7 &  
  $0.2431877-0.0832337i $&$ 0.2432030 - 0.0832420 i$    &  0.0067756  &
  $0.2194009-0.0763619i $&$ 0.219372- 0.0764423i$    & 0.036808\\  \hline
  \multirow{3}{*}{-1}
 & 0.3 &  
 $0.211852 -0.0776963i $&$0.21188-0.0777303  i$    & 0.019446   &
 $0.191904-0.0707087i $&$0.191882 - 0.0707742  i$    & 0.033769 \\ 
  & 0.5 &  
 $0.22384-0.0797129i$ &$0.22384-0.0797129 i$   &  0.018775   &
 $0.202397-0.0727402i$ &$0.20241-0.0727948 i$   &   0.026132 \\
   & 0.7 &  
 $0.244093-0.0817178i $&$0.244256-0.0815949  i$    &  0.079390   &
 $0.220598-0.0749383i $&$0.22087-0.0747863  i$    &  0.13340  \\  \hline
\end{tabular}}
\end{table}

Tables. \ref{S0} and \ref{E1}  present the fundamental quasinormal modes ($n=0$) calculated using different numerical methods. These tables not only list the frequencies but also provide a rigorous accuracy comparison through the relative errors $\varepsilon_1$ (between AIM and Bernstein methods) and $\varepsilon_2$ (between AIM and WKB methods). From Table. \ref{S0}, which focuses on the scalar perturbation with $l=0$, we observe that $\varepsilon_1$ is extremely small (typically less than 0.5\%), indicating excellent agreement between the AIM and the Bernstein spectral method. This cross-validation confirms the high precision of our results for the $l=0$ mode. In contrast, the relative error for the WKB method, $\varepsilon_2$, is noticeably larger (reaching up to 9\%). This quantitative evidence supports our earlier assertion that the WKB approximation is less reliable for low angular momentum modes ($l=0$) due to the double-peak structure of the effective potential, thereby necessitating the use of the Bernstein method for this sector. Table. \ref{E1} displays the results for higher multipole numbers ($l=1$) for both scalar and electromagnetic perturbations. Here, the relative error $\varepsilon_2$ remains consistently negligible (mostly below 0.1\%). This result demonstrates that for $l\geq 1$, the WKB approximation converges significantly and provides highly accurate frequencies comparable to the AIM. Consequently, these comparisons validate the numerical robustness of the hybrid approach employed in this work.

\subsection{$\Lambda$-dependence}

\begin{figure}[H]
\centering
\subfigure[]
{%% label for first subfigure
\includegraphics[width=2.5in]{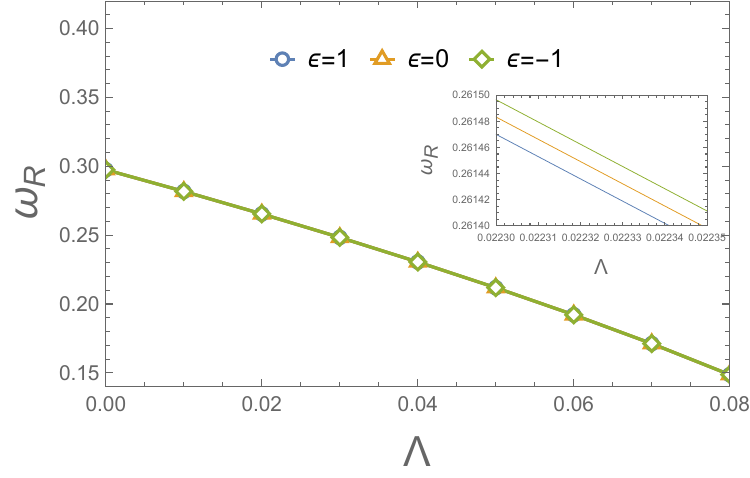}}
\subfigure[]
{%% label for first subfigure
\includegraphics[width=2.55in]{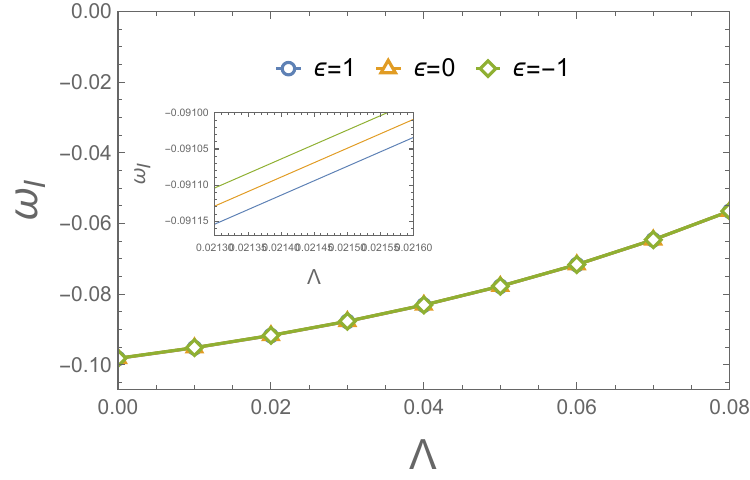}}
\caption{Variation of scalar fundamental QNFs, ($n=0$) with respect to the cosmological constant $\Lambda$ for black hole with fixed $M=1$, $Q_{\mathrm{m}}=0.3$ and $l=1$. Panels (a) and (b) display $\text{Re}(\omega)$ and $|\text{Im}(\omega)|$, respectively. The curves correspond to different coupling parameters: $\epsilon=1$ (blue), $\epsilon=0$ (orange), and $\epsilon=-1$ (green). As $\Lambda$ increases, both $\text{Re}(\omega)$ and $|\text{Im}(\omega)|$ decrease monotonically.}\label{figQ2}
\end{figure}

\begin{figure}[H]
\centering
\subfigure[]
{%% label for first subfigure
\includegraphics[width=2.5in]{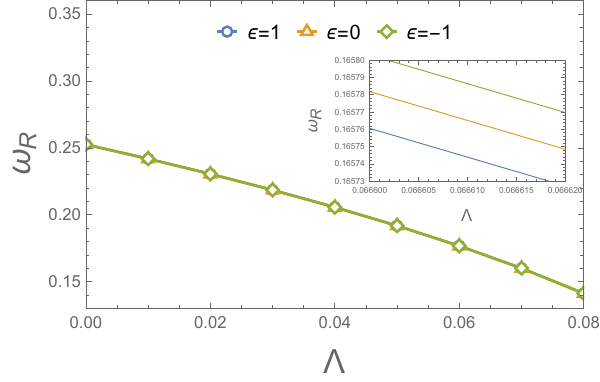}}
\subfigure[]
{%% label for first subfigure
\includegraphics[width=2.55in]{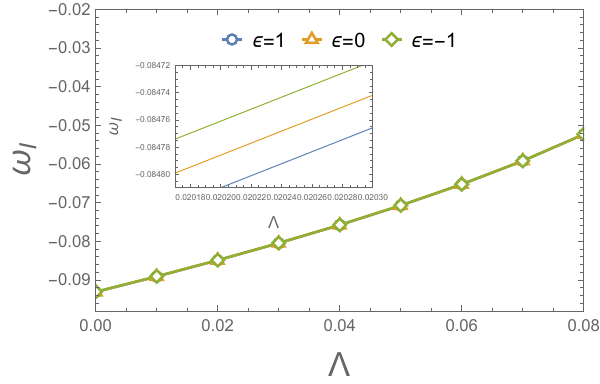}}
\caption{Variation of electromagnetic fundamental QNFs, ($n=0$) with respect to the cosmological constant $\Lambda$ for black hole with fixed $M=1$,$Q_{\mathrm{m}}=0.3$ and $l=1$. Panels (a) and (b) display $\text{Re}(\omega)$ and $|\text{Im}(\omega)|$, respectively. The curves correspond to different coupling parameters: $\epsilon=1$ (blue), $\epsilon=0$ (orange), and $\epsilon=-1$ (green). As $\Lambda$ increases, both $\text{Re}(\omega)$ and $|\text{Im}(\omega)|$ decrease monotonically.}\label{figQ20}
\end{figure}

In this subsection, we investigate the influence of the cosmological constant $\Lambda$ on the quasinormal modes for both scalar and electromagnetic perturbations. The results are detailed in Table \ref{sl1}. In these calculations, the black hole mass and magnetic charge are fixed at $M=1, Q_{\mathrm{m}}=0.3$ respectively. We also plot the  the real part (oscillation frequency) and the imaginary part
(damping rate) of the fundamental QNFs (n = 0) for the scalar and  electromagnetic perturbations in Figs.\ref{figQ2} and \ref{figQ20}. A clear trend emerges from the data and figures: as the cosmological constant $\Lambda$ increases, both the real part of the QNF $\text{Re}(\omega)$ and the magnitude of the imaginary part $|\text{Im}(\omega)|$ decrease monotonically for both types of perturbations. Physically, the real part $\text{Re}(\omega)$ corresponds to the oscillation frequency of the perturbation, while the imaginary part $|\text{Im}(\omega)|$ represents the damping rate. Therefore, a larger cosmological constant leads to perturbations that oscillate at a lower frequency and decay more slowly (i.e., are longer-lived). This can be understood by considering that a larger $\Lambda$ reduces the size of the potential cavity between the black hole and cosmological horizons, which in turn supports lower-frequency, longer-damped resonant modes. The range of the cosmological constant $\Lambda$ is restricted to $\Lambda\in[0,0.08]$, which is well below the critical value $\Lambda_{\mathrm{c}}\approx 0.11$ where the black hole event horizon $r_{\mathrm{h}}$ and the cosmological horizon $r_{\mathrm{c}}$ coincide. This limitation is essential because, as $\Lambda$ approaches $\Lambda_{\mathrm{c}}$ (in Figs.\ref{fig33}), the potential barrier separating the two horizons becomes extremely narrow, leading to significant numerical instability and rendering the calculated QNMs unreliable in that region.

\begin{table*}[h!]
\caption{The fundamental QNFs ($n=0$) of scalar perturbations and electromagnetic perturbations $l=1$ 
for various $\epsilon$ and $\Lambda$. Other parameters are set to $M=1$ and $Q_{\mathrm{m}}=0.3$. The parameter $\varepsilon_2$ represents relative errors between AIM and WKB methods.}\label{sl1}
\resizebox{\linewidth}{!}{
\begin{tabular}{|c|c|c|c|c|c|c|c|} \hline
&  & \multicolumn{3}{|c|}{scalar field perturbation} & \multicolumn{3}{|c|}{electromagnetic field perturbation}   \\ \hline
$\epsilon$&  $\Lambda$ &  AIM  &  WKB  &   $\varepsilon_2(\%)$   &  AIM  &  WKB  &   $\varepsilon_2(\%)$            \\ \hline
\multirow{3}{*}{1}& 0.01  & $0.281768 - 0.0951460 i$  & $0.281735 - 0.0952336 i$       &  0.031390    & $0.241923 - 0.0891115  i$  & $0.24182 - 0.0892346 i$       &  0.062358        \\ 
 & 0.05 &$0.211816 - 0.0777350i$ & $0.211845 - 0.0777674  i$   & 0.019121& 
 $0.191863 - 0.0707426i$ & $0.191835 - 0.0708107  i$   &0.036163    \\ 
  & 0.07 &$0.171217 - 0.0645797 i$  & $0.171236 - 0.0646951i$   &0.063913 & 
  $0.160089 - 0.0591985 i$  & $0.160138 - 0.059216i$   &0.030115    \\ \hline
  \multirow{3}{*}{0}& 0.01  & $0.281779 - 0.0951194i$  & $0.281753 - 0.095204i$      & 0.029778  & $0.241944 - 0.0890852i$  & $0.241854 - 0.0892002 i$      &  0.056664    \\ 
   & 0.05 &$0.211834 - 0.0777157i$ & $0.211861 - 0.0777493   i$    & 0.019143& 
   $0.191884 - 0.0707257i$ & $0.191852 - 0.0707951   i$    & 0.037424\\ 
   & 0.07 &$0.171239 - 0.0645665 i$  & $0.171249 - 0.0646861i$   &0.065575& 
   $0.160111 - 0.0591870i$  & $0.160065 - 0.05924i$   &0.041105 \\\hline
   \multirow{3}{*}{-1}& 0.01  & $0.281791 - 0.0950926 i$  & $0.281771 - 0.0951741 i$       & 0.028222  & $0.241966 - 0.0890587 i$  & $0.24189 - 0.0891654i$       & 0.050760      \\ 
   & 0.05 &$0.211852 - 0.0776963 i$ & $0.21188 - 0.0777303 i$    & 0.019450&
   $0.191904 - 0.0707087i$ & $0.191879 - 0.0707755   i$    &0.034891\\ 
   & 0.07 &$0.171260 - 0.0645533$   & $0.171295 - 0.0646648 i$   &0.063687&
   $0.160132 - 0.0591754i$  & $0.160085 - 0.0592292i$   &0.041846\\\hline
 \end{tabular}}
\end{table*}

\subsection{$\epsilon$-dependence}

\begin{figure}[H]
\centering
\subfigure[]
{%% label for first subfigure
\includegraphics[width=2.5in]{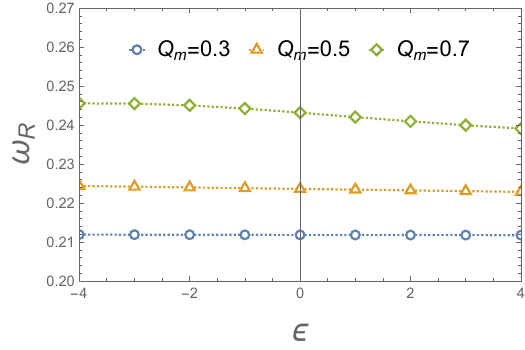}}
\subfigure[]
{%% label for first subfigure
\includegraphics[width=2.55in]{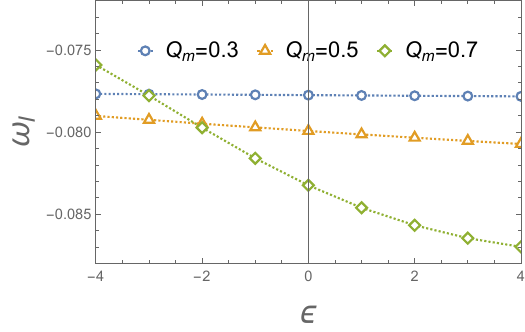}}
\caption{Variation of scalar fundamental QNFs ($n=0$) with respect to the nonlinear coupling parameter $\epsilon$ for black hole with fixed $M=1$, $\Lambda=0.05$ and $l=1$. Panel (a) displays $\text{Re}(\omega)$ and panel (b) shows $\text{Im}(\omega)$. Each panel compares results for three different magnetic charges: $Q_{\mathrm{m}}=0.3$ (blue), $Q_{\mathrm{m}}=0.5$ (orange), and $Q_{\mathrm{m}}=0.7$ (green). For smaller magnetic charges ($Q_{\mathrm{m}}=0.3, 0.5$), both $\text{Re}(\omega)$ and $|\text{Im}(\omega)|$ exhibit minimal dependence on $\epsilon$. However, for the larger magnetic charge ($Q_{\mathrm{m}}=0.7$), increasing $\epsilon$ leads to a moderate decrease in both oscillation frequency and damping rate.}\label{figep2}
\end{figure}

\begin{figure}[H]
\centering
\subfigure[]
{%% label for first subfigure
\includegraphics[width=2.5in]{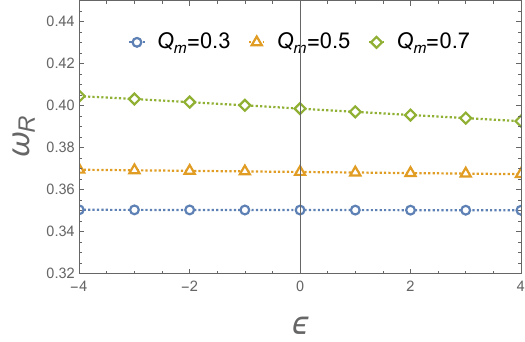}}
\subfigure[]
{%% label for first subfigure
\includegraphics[width=2.55in]{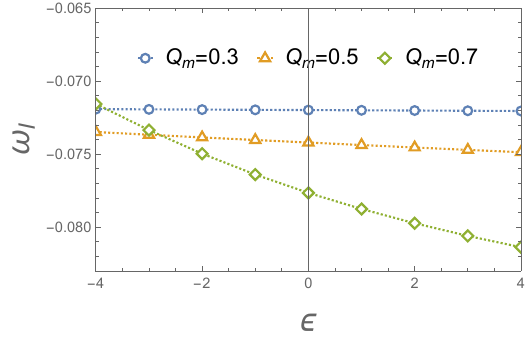}}
\caption{Variation of electromagnetic fundamental QNFs ($n=0$) with respect to the nonlinear coupling parameter $\epsilon$ for black hole with fixed $M=1$, $\Lambda=0.05$ and $l=1$. Panel (a) displays $\text{Re}(\omega)$ and panel (b) shows $\text{Im}(\omega)$. Each panel compares results for three different magnetic charges: $Q_{\mathrm{m}}=0.3$ (blue), $Q_{\mathrm{m}}=0.5$ (orange), and $Q_{\mathrm{m}}=0.7$ (green). For smaller magnetic charges ($Q_{\mathrm{m}}=0.3, 0.5$), both $\text{Re}(\omega)$ and $|\text{Im}(\omega)|$ exhibit minimal dependence on $\epsilon$. However, for the larger magnetic charge ($Q_{\mathrm{m}}=0.7$), increasing $\epsilon$ leads to a moderate decrease in both oscillation frequency and damping rate.}\label{figep20}
\end{figure}

Finally, we analyze the effect of the nonlinear electromagnetic coupling parameter $\epsilon$ on the quasinormal modes. The variation of the fundamental QNFs for scalar and electromagnetic perturbations as a function of $\epsilon$ is displayed in Figs.\ref{figep2} and \ref{figep20}, respectively. The most striking result is that the parameter $\epsilon$ has a remarkably weak influence on the fundamental QNFs. As shown in the two figures, both the real and imaginary parts of the QNFs remain almost constant even as $\epsilon$ varies over a wide range when the magnetic charge fixed $Q_{\mathrm{m}}=0.3$ and $0.5$. When $Q_{\mathrm{m}}=0.7$, the real or imaginary part of the QNFs shows a more noticeable change as $\epsilon$ varies. These behaviors correspond to the potential function graphs, see Figs.\ref{fig60}.

\section{ The Greybody factor}
\label{sec6}

In this section, we use the WKB approximation method to calculate greybody factors for scalar and  ectromagnetic perturbation. 
The boundary condition for the scattering process is different from that of the QNMs, which can be written as
\begin{eqnarray}
\psi& =& T(\omega)\mathrm{e}^{-\mathrm{i}\omega r_*}, \quad r_* \rightarrow -\infty\nonumber,\\
\psi& =& \mathrm{e}^{-\mathrm{i}\omega r_*} + R(\omega)\mathrm{e}^{\mathrm{i}\omega r_*}, \quad r_* \rightarrow +\infty,\label{boundgrey}
\end{eqnarray}
where $R$ and $T$ represent the reflection coefficient and transmission coefficient, respectively. The greybody factor is defined as the probability of an outgoing wave
reaching to infinity or an incoming wave absorbed by the black hole. Therefore, $|T(\omega)|^2$ is called the greybody factor, and $R(\omega)$ and $T(\omega)$ should satisfy the following relation
\begin{eqnarray}
|R(\omega)|^2+|T(\omega)|^2=1.
\end{eqnarray}

Using the 6th-order WKB method, the
reflection and transmission coefficients can be obtained
\begin{eqnarray}
&&|R(\omega)|^2 = \frac{1} {1 + \mathrm{e}^{-2\pi \mathrm{i} \mathcal{K}(\omega)}} ,\nonumber\\
&&|T(\omega)|^2 =\frac{1} {1 + \mathrm{e}^{2\pi \mathrm{i} \mathcal{K}(\omega)}}= 1 -|R(\omega)|^2,\label{TR}
\end{eqnarray}
where $\mathcal{K}$ is a parameter which can be obtained by the WKB formula
\begin{eqnarray}
\mathcal{K}= \frac{\mathrm{i}\left( \omega^2 - V(r_0) \right)}{\sqrt{-2V''(r_0)}} + \sum_{i=2}^6 \Lambda_i.
\end{eqnarray}

Notice that the accuracy of the WKB approximation is, in fact, excellent for any \( n < l \), and for a given $n$, the relative agreement improves with increasing $l$ \cite{Iyer:1986nq,Konoplya:2003ii,Liu:2023kxd}. 

\begin{figure}[htb]
\centering
\subfigure[$\epsilon=-1$  ]
{\label{FSQ1} %% label for first subfigure
\includegraphics[width=2in]{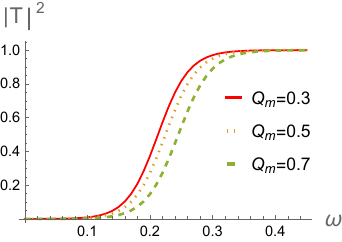}}
\hfill
\subfigure[$\epsilon=0$]
{\label{FSQ2} %% label for first subfigure
\includegraphics[width=2in]{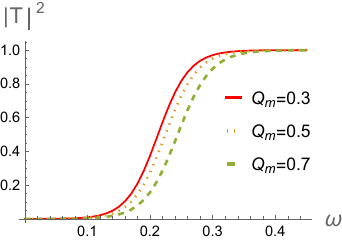}}
\hfill
\subfigure[ $\epsilon=1$]
{\label{FSQ3} %% label for first subfigure
\includegraphics[width=2in]{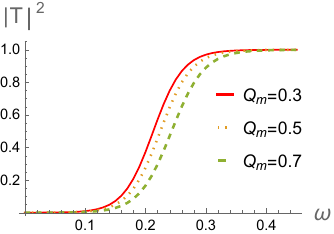}}
\hfill
\caption{Greybody factors $\|T(\omega)\|^2$ for massless scalar field perturbations on black hole with fixed $M=1$, $\Lambda=0.05$, and $l=1$. The three panels display results for different coupling parameters $\epsilon=-1, 0, +1$. Each panel compares the transmission probabilities for three values of magnetic charge $Q_{\mathrm{m}}=0.3$ (red), $Q_{\mathrm{m}}=0.5$ (orange), and $Q_{\mathrm{m}}=0.7$ (green). Notably, the greybody factor decreases monotonically with increasing magnetic charge $Q_{\mathrm{m}}$ for all coupling parameters.}\label{FSQ}
\end{figure}

\begin{figure}[htb]
\centering
\subfigure[$\epsilon=-1$  ]
{\label{FEQ1} %% label for first subfigure
\includegraphics[width=2in]{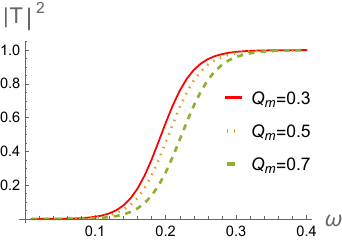}}
\hfill
\subfigure[$\epsilon=0$]
{\label{FEQ2} %% label for first subfigure
\includegraphics[width=2in]{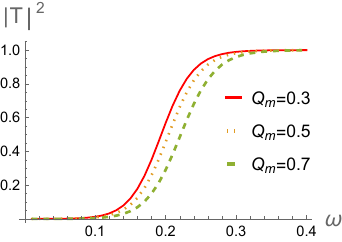}}
\hfill
\subfigure[ $\epsilon=1$]
{\label{FEQ3} %% label for first subfigure
\includegraphics[width=2in]{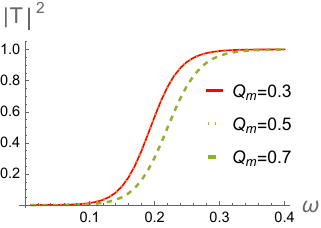}}
\hfill
\caption{Greybody factors $\|T(\omega)\|^2$ for electromagnetic field perturbations on black hole with fixed $M=1$, $\Lambda=0.05$, and $l=1$. The three panels display results for different coupling parameters $\epsilon=-1, 0, +1$. Each panel compares the transmission probabilities for three values of magnetic charge $Q_{\mathrm{m}}=0.3$ (red), $Q_{\mathrm{m}}=0.5$ (orange), and $Q_{\mathrm{m}}=0.7$ (green). Notably, the greybody factor decreases monotonically with increasing magnetic charge $Q_{\mathrm{m}}$ for all coupling parameters.}\label{FEQ}
\end{figure}

\begin{figure}[htb]
\centering
\subfigure[$\epsilon=-1$  ]
{\label{FSL1} %% label for first subfigure
\includegraphics[width=2in]{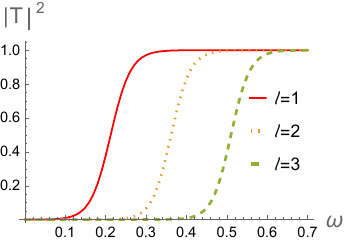}}
\hfill
\subfigure[$\epsilon=0$]
{\label{FSL2} %% label for first subfigure
\includegraphics[width=2in]{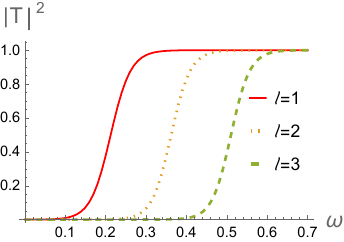}}
\hfill
\subfigure[ $\epsilon=1$]
{\label{FSL3} %% label for first subfigure
\includegraphics[width=2in]{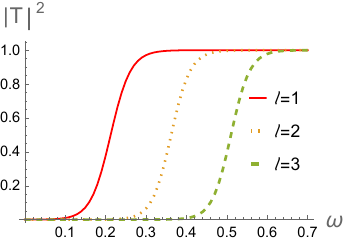}}
\hfill
\caption{Greybody factors $\|T(\omega)\|^2$ for massless scalar field perturbations on black hole with fixed $M=1$, $\Lambda=0.05$, and $Q_{\mathrm{m}}=0.3$. The three panels display results for different coupling parameters $\epsilon=-1, 0, +1$. Each panel compares the transmission probabilities for three values of $l=1$ (red), $l=2$ (orange), and $l=3$ (green). Notably, the greybody factor decreases monotonically with increasing $l$ for all coupling parameters.}\label{FSL}
\end{figure}

\begin{figure}[htb]
\centering
\subfigure[$\epsilon=-1$  ]
{\label{FEL1} %% label for first subfigure
\includegraphics[width=2in]{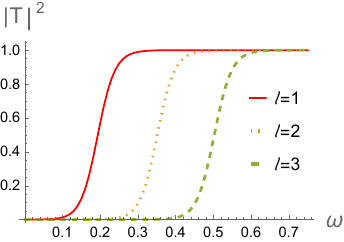}}
\hfill
\subfigure[$\epsilon=0$]
{\label{FEL2} %% label for first subfigure
\includegraphics[width=2in]{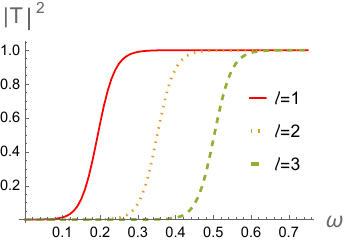}}
\hfill
\subfigure[ $\epsilon=1$]
{\label{FEL3} %% label for first subfigure
\includegraphics[width=2in]{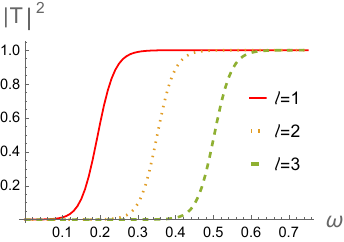}}
\hfill
\caption{Greybody factors $\|T(\omega)\|^2$ for electromagnetic field perturbations on black hole with fixed $M=1$, $\Lambda=0.05$, and $Q_{\mathrm{m}}=0.3$. The three panels display results for different coupling parameters $\epsilon=-1, 0, +1$. Each panel compares the transmission probabilities for three values of $l=1$ (red), $l=2$ (orange), and $l=3$ (green). Notably, the greybody factor decreases monotonically with increasing $l$ for all coupling parameters.}\label{FELQ}
\end{figure}

\begin{figure}[htb]
\centering
\subfigure[$Q_{\mathrm{m}}=0.3$  ]
{\label{FS1} %% label for first subfigure
\includegraphics[width=2in]{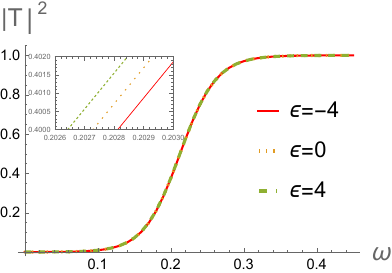}}
\hfill
\subfigure[$Q_{\mathrm{m}}=0.5$]
{\label{FS2} %% label for first subfigure
\includegraphics[width=2in]{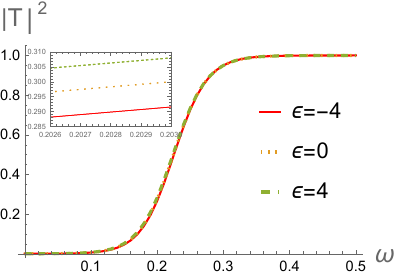}}
\hfill
\subfigure[ $Q_{\mathrm{m}}=0.7$]
{\label{FS3} %% label for first subfigure
\includegraphics[width=2in]{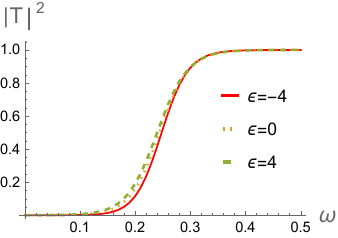}}
\hfill
\caption{Greybody factors $\|T(\omega)\|^2$ for massless scalar field perturbations on black hole with fixed $M=1$, $\Lambda=0.05$, and $l=1$. The three panels display results for different magnetic charges $Q_{\mathrm{m}}=0.3, 0.5, 0.7$. Each panel compares the transmission probabilities for three values of coupling parameters $\epsilon=-4$ (red), $\epsilon=0$ (orange), and $\epsilon=4$ (green). Notably, the greybody factor increases monotonically with increasing $\epsilon$ for all magnetic charges.}\label{FS}
\end{figure}

\begin{figure}[htb]
\centering
\subfigure[$Q_{\mathrm{m}}=0.3$  ]
{\label{FE1} %% label for first subfigure
\includegraphics[width=2in]{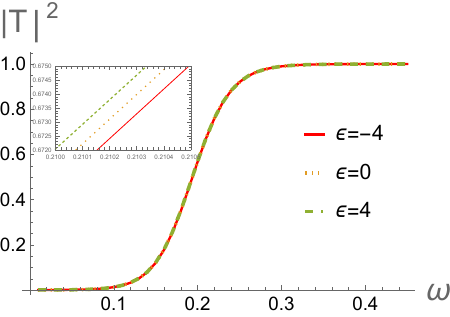}}
\hfill
\subfigure[$Q_{\mathrm{m}}=0.5$]
{\label{FE2} %% label for first subfigure
\includegraphics[width=2in]{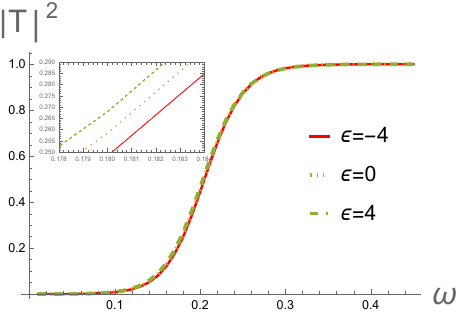}}
\hfill
\subfigure[ $Q_{\mathrm{m}}=0.7$]
{\label{FE3} %% label for first subfigure
\includegraphics[width=2in]{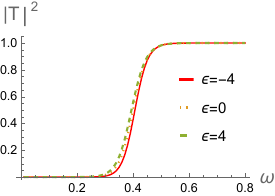}}
\hfill
\caption{Greybody factors $\|T(\omega)\|^2$ for electromagnetic field perturbations on black hole with fixed $M=1$, $\Lambda=0.05$, and $l=1$. The three panels display results for different magnetic charges $Q_{\mathrm{m}}=0.3, 0.5, 0.7$. Each panel compares the transmission probabilities for three values of coupling parameters $\epsilon=-4$ (red), $\epsilon=0$ (orange), and $\epsilon=4$ (green). Notably, the greybody factor increases monotonically with increasing $\epsilon$ for all magnetic charges.}\label{FE}
\end{figure}

In Figs.~\ref{FSQ}-\ref{FE}, we show the behaviors of the greybody factors for scalar and  ectromagnetic perturbation under varying the magnetic charge $Q_{\mathrm{m}}$, multipole number $l$ and parameter $\epsilon$. It's seen that the greybody factors exhibit noticeable decrease as $Q_{\mathrm{m}}$ increases (see Fig.~\ref{FSQ}-\ref{FEQ}). This
suggests that when the magnetic charge $Q_{\mathrm{m}}$ is reduced, the black hole becomes more effective at capturing and interacting with incoming matter or radiation. We also investigate how a change of angular number $l$ in affects the corresponding behavior of the greybody factors (see Fig.~\ref{FSL}-\ref{FELQ}). The greybody factors also gradually decrease as $l$ increases, which reveals that the greybody factors are higher for smaller values of $l$. With regard to difference $\epsilon$, the greybody factors gradually increase as $\epsilon$ increases (see Fig.~\ref{FS}-\ref{FE}). It indicates that black holes become less interactive with the surrounding radiation and allow more of perturbed field to escape. These are all consistent with the effective potentials in Figs.\ref{fig2}-\ref{fig60}.


\section{Conclusion and discussion}
\label{sec7}

In this paper, we have discussed the perturbations of different test fields on magnetically charged de Sitter (dS) black holes in string-inspired Euler-Heisenberg theory, which involves a scalar field $\phi$ coupled to the electromagnetic field via a non-linear function $f(\phi)$. Moreover, this black hole exhibits different horizon structures for $\epsilon>0$ and $\epsilon<0$. The influences of magnetic charge $Q_{\mathrm{m}}$, the parameter $\epsilon$ and the angular quantum number $l$ on the corresponding effective potentials of the perturbations have been analyzed.
We then accurately calculated the Quasinormal Frequencies (QNFs) for massless scalar and electromagnetic field perturbations using the Asymptotic Iteration Method (AIM) and the WKB approximation, with the results from both techniques showing excellent agreement. Crucially, to address the potential failure of the WKB approximation for $l=0$ scalar field perturbations, we employed the Bernstein spectral method as a verification tool, further solidifying the robustness of our results.

The analysis of primary parameters clearly demonstrated the accelerating effect of the magnetic charge $Q_{\mathrm{m}}$: an increase in $Q_{\mathrm{m}}$ leads to a monotonic increase in both the real part (oscillation frequency) and the absolute value of the imaginary part (damping rate) of the QNFs. This suggests that black holes with larger magnetic charges oscillate at higher frequencies and return to equilibrium more quickly. In sharp contrast to the effect of $Q_{\mathrm{m}}$, the cosmological constant $\Lambda$, which defines the dS background, exhibits a noticeable inhibitory effect on the black hole's perturbation dynamics. As $\Lambda$ increases, both the real part and the absolute value of the imaginary part of the QNFs decrease monotonically. Physically, a larger $\Lambda$ corresponds to a smaller cosmological horizon, which effectively shortens the "round-trip" distance for the perturbation wave between the event horizon and the cosmological horizon. This results in a slower energy dissipation and consequently a slower decay rate. This $\Lambda$-driven damping effect stands in direct opposition to the accelerating effect caused by the magnetic charge $Q_{\mathrm{m}}$, with both parameters collectively shaping the final quasi-normal spectrum. 

In summary, our work not only confirms the structural stability of magnetically charged dS black holes in EH gravity but also quantifies the combined influences of the three critical parameters—magnetic charge $Q_{\mathrm{m}}$ non-linear coupling $\epsilon$, and the cosmological background $\Lambda$—on the black hole's quasi-normal oscillations. These results provide essential theoretical benchmarks for future gravitational wave astronomy and the testing of modified gravity theories. The calculated greybody factors serve as a direct link between the black hole's dynamical perturbation and its observable Hawking radiation spectrum, laying the groundwork for potentially extracting evidence of EH theory and the dS cosmic background information from astrophysical data.

 \vspace{1cm}
{\bf Acknowledgments}
 \vspace{0.5cm}

We gratefully acknowledge support by the National Natural Science Foundation of China (NNSFC) (Grant No.12365009), Jiangxi Provincial Natural Science Foundation (Grant No. 20232BAB201039) and Natural Science Basic Research Program of Shaanxi Province (Program No.2023-JC-QN-0053) and (Program No.2023-JC-QN-0267). 

\bibliographystyle{unsrt}
\bibliography{reference}

\end{document}